\documentclass[range]{ar2e}

\usepackage{graphicx}
\usepackage{bm}   
\usepackage{epsf}
\usepackage{rotating}
\usepackage{epsfig,graphics,rotate,color}
\usepackage{wrapfig}
\usepackage{amssymb}
\usepackage{amsmath}
\usepackage{amsfonts}

\newcommand{\gtwid}{\mathrel{\raise.3ex\hbox{$>$\kern-.75em\lower1ex\hbox{$\sim$}}}}
\newcommand{\ltwid}{\mathrel{\raise.3ex\hbox{$<$\kern-.75em\lower1ex\hbox{$\sim$}}}}

\begin{document}



\title{The LSND and MiniBooNE Oscillation Searches at High $\Delta m^2$}

\author{Janet~M.~Conrad
\affiliation{Massachusetts Institute of Technology; Cambridge, MA 02139}
William~C.~Louis
\affiliation{Los Alamos National Laboratory; Los Alamos, NM 87545}
Michael~H.~Shaevitz
\affiliation{Columbia University; New York, NY 10027}}

\begin{abstract}

This paper reviews the results of the LSND and MiniBooNE experiments.
The primary goal of each experiment was to effect sensitive searches
for neutrino oscillations in the mass region with $\Delta m^2 \sim 1$ eV$^2$.  The two
experiments are complementary, and so the comparison of results can bring additional
information with respect to models with sterile neutrinos.   Both experiments obtained evidence for
$\bar \nu_\mu \rightarrow \bar \nu_e$ oscillations, and MiniBooNE also 
observed a $\nu_\mu \rightarrow \nu_e$  excess.
In this paper, we review the design, analysis, and results from these experiments.   We then
consider the results within the global context of sterile neutrino
oscillation models.   The final data sets require a more extended model than
the simple single sterile neutrino model 
imagined at the time that LSND drew to a close and MiniBooNE began.    
We show that there are apparent incompatibilities between
data sets in models with two sterile neutrinos.   However, these
incompatibilities
may be explained with variations within the systematic error. Overall,
models with two (or three) sterile neutrinos seem to succeed
in fitting the global data, and they
make interesting predictions for future experiments.

\end{abstract}

\maketitle

\section{Introduction \label{intro}}

In the past 20 years, neutrino oscillations--where a beam of one flavor of
neutrino evolves to have components of other flavors--have gone from speculation 
to demonstration.  This has motivated an extension to particle physics 
phenomenology that includes finite neutrino masses and
a $(3\times3)$ mixing matrix that connects 
the three known neutrino flavors, $\nu_e$, $\nu_\mu$, and
$\nu_\tau$,  to the three mass states, $\nu_1$, $\nu_2$, and $\nu_3$ \cite{1209.3023}.
However, this extension may not be the entire story.  Indeed,
experimental clues are arising that seem to point to additional, noninteracting
(``sterile'') neutrino states.    Among the experimental results, 
the LSND and MiniBooNE  experiments have been crucial motivators for the  
recent explosion of interest in this topic \cite{whitepaper}.   This
excitement in the community, 
as well as the end of MiniBooNE running in 2012, 
provides the {\it raison d'etre} for a review of the  LSND and MiniBooNE
results at this time.

To understand why richer phenomenology is motivated, one 
must consider the observed ``mass splittings'' measured in neutrino oscillation
experiments.  These are the differences 
between the squared mass values.  A model
with three neutrinos will have two independent mass splittings.   
In the presently accepted picture of neutrino oscillations,
the magnitude of the mass splitting between $\nu_1$ and $\nu_2$
is
$\Delta m^2_{21} =(7.50 \pm0.185)\times 10^{-5}$ eV$^2$,  
and $\nu_2$ and $\nu_3$ is
$\Delta m^2_{32}=(2.47^{+0.069}_{-0.067}) \times
10^{-3}$ eV$^2$ \cite{1209.3023}.
However, experimental results indicate $>3\sigma$ evidence for oscillations 
with $\Delta m^2 \sim 1$ eV$^2$. 
A way to accommodate this higher mass splitting is the 
introduction of sterile neutrinos that act as new flavors that
mix with the Standard Model flavors but do not couple to the $W$
or $Z$, thereby avoiding the limits on extra active neutrino flavors \cite{Zdecay}.

This paper focuses primarily on LSND and MiniBooNE,  two experiments 
that were  constructed to probe the $\Delta m^2\sim$1 eV$^2$ region.
In 1995, LSND was the first experiment to
publish evidence for a signal at $\Delta m^2 \sim 1$ eV$^2$,
observing $\bar \nu_\mu \rightarrow \bar \nu_e$ 
\cite{lsndPRL95, lsndPRL96, lsndPRL98, lsndPRD}.  
MiniBooNE was proposed \cite{MBprop}  in 1998 to follow up
on this signal, while substantially changing the systematics.
MiniBooNE saw excesses in $\nu_\mu \rightarrow \nu_e$ and
$\bar \nu_\mu \rightarrow \bar \nu_e$ modes \cite{MBnu, MBnubar09, MBnubar}.  

This paper examines the LSND and MiniBooNE appearance signals.
We begin, in Sec.~\ref{pheno}, by discussing the oscillation phenomenology that will be used
throughout the paper.    
In Sec.~\ref{expts} we introduce the LSND
and MiniBooNE experiments, drawing important distinctions between the designs.  
The primary results of the experiments,  the appearance signals, are presented 
in Sec.~\ref{results}.    LSND and MiniBooNE have also reported 
$\nu_e$ and $\nu_\mu$ disappearance limits, and other experiments 
sensitive to the $\Delta m^2 \sim 1$ eV$^2$ region have presented
relevant results as well.
The LSND and MiniBooNE appearance signal cannot be
interpreted outside of consideration of these other data sets, which
we report  
in Sec.~\ref{global}.   Then, in Sec.~\ref{sterilecon}, we end by using those data sets,   
to explore potential  
sterile neutrino models that may explain the LSND and MiniBooNE results.   We make the case that models
with at least two additional sterile states, and that include $CP$
violation, can successfully describe all of the data
sets.   These models are testable in the near future.

\section{Phenomenology of Oscillations Involving Sterile Neutrinos \label{pheno}}

We will interpret LSND, MiniBooNE, and other experimental results within neutrino oscillation models.
Oscillations arise if neutrinos are produced, through weak decay, in a specific flavor
eigenstate that is a combination of mass eigenstates.   The mass
eigenstates propagate with slightly different frequencies, leading to
neutrino-flavor probability waves with oscillating beats among the flavors. 

In a simple two-neutrino model, 
the flavor states, $\nu_e$ and $\nu_\mu$,
are linked to the mass states, $\nu_1$ and $\nu_2$, via a mixing matrix
that is a simple rotation matrix.  At $t=0$, the neutrinos are described by:
\begin{eqnarray}
|\nu _e \rangle =\cos \theta \; |\nu _1\rangle+\sin \theta \;|\nu
_2 \rangle \ {\rm and}\\ 
|\nu _\mu \rangle =-\sin \theta \;|\nu _1\rangle+\cos \theta \;|\nu
_2 \rangle,
\end{eqnarray}
where $\theta$ is called the ``mixing angle.''   
The quantum mechanical interference of the mass eigenstates as they propagate  leads to 
the appearance oscillation probability:
\begin{equation}
P=\sin^2
2\theta \ \sin^2(1.27 \Delta m^2(L/E)),
\label{osceq}
\end{equation}
where $\theta$ is the mixing angle, $L$, in m, is the distance from production to
detection, 
$E$, in MeV, is the energy of the neutrino, and
$\Delta m^2 =
m_{2}^2-m_{1}^2$, in eV$^2$, is the mass splitting.
The disappearance probability is given by
\begin{equation}
P =1-\sin^2
2\theta \ \sin^2(1.27 \Delta m^2(L/E)). 
\label{disosceq}
\end{equation}

From the above, one sees that for LSND and MiniBooNE to be sensitive to 
$\Delta m^2 \sim 1$~eV$^2$, one chooses $L/E \sim 1$ m/MeV  
to maximize the oscillation probability.    Because typical beam
energies range from a few MeV to a few GeV, this $L/E$ demands
experiments that are relatively close to the site of neutrino
production.
As a result, these are called ``short baseline'' (SBL) experiments.

Any single SBL experiment is likely 
to fit well within this simple two-neutrino model, producing either a
signal or a limit at a given confidence level (CL).  However, adding additional data sets 
generally requires extensions to more than two flavors. 
The phenomenological extension to three active neutrinos provides
an example of how to enlarge the model \cite{1209.3023}.    
 In a series of recent papers, we have explored how to take the next step
of also introducing sterile netrinos states\cite{sorel, georgiaCP,
  georgiaVaiability}.
Reference~\cite{Ignarra} provides a good step-by-step review of
the ideas which we summarize briefly here.

The simplest extension is the ``3+1 model,'' which introduces one 
sterile neutrino, labeled ``s,'' that can mix with the three active
flavors.  This leads to a mixing matrix that connects the 3+1 flavors
to four mass states:
\[
\left( 
\begin{array}{l}
\nu _e \\ 
\nu _\mu \\ 
\nu _\tau \\
\nu_s
\end{array}
\right) =\left( 
\begin{array}{llll}
U_{e1} & U_{e2} & U_{e3} & U_{e4} \\ 
U_{\mu 1} & U_{\mu 2} & U_{\mu 3} & U_{\mu 4} \\ 
U_{\tau 1} & U_{\tau 2} & U_{\tau 3} & U_{\tau 4} \\
U_{s 1} & U_{s 2} & U_{s 3} & U_{s 4}
\end{array}
\right) \left( 
\begin{array}{l}
\nu _1 \\ 
\nu _2 \\ 
\nu _3 \\
\nu _4 
\end{array}
\right)~. \label{mx}
\]
The mixing among the three active flavor states is highly constrained by
measurements from Daya Bay \cite{DB},
Double Chooz \cite{DCGd}, KamLAND \cite{KL}, MINOS \cite{MINOS, MINOSdis}, 
RENO \cite{RENO},  Super K \cite{SK}, SNO \cite{SNOfinal}, and T2K
\cite{T2K, T2Kdis}.   Accommodating these data requires 
that three of the mass states be mostly active flavors,  which leads
to the fourth mass state,  $\nu_4$, being primarily
sterile.  This model assumes that $|U_{\tau 4}|$ is negligible, for simplicity.
The ``SBL approximation,''  
$\Delta m^2_{21} \approx \Delta m^2_{31} \equiv 0$, is applied, motivated
by the assumption that the  $\nu_4$ mass is much larger than the other mass
states.     As a result, 
the $\nu_\mu \rightarrow \nu_e$ appearance probability 
simplifies to 
Eq.~\ref{osceq}, with $\sin^2 2\theta$ given by
\begin{equation}
\sin^22\theta_{e \mu}  =  4 U_{e4}^2U_{\mu 4}^2. \label{angleapp}
\end{equation}   
The disappearance probabilities are given by Eq.~\ref{disosceq} with
\begin{eqnarray}
\sin^22\theta_{\mu \mu} & = & 4 U_{\mu 4}^2 (1-U_{\mu 4}^2) ~~~(\mu~flavor), \label{anglemumu} \\
\sin^22\theta_{e e} & = & 4 U_{e 4}^2 (1-U_{e 4}^2) ~~~(e~flavor). \label{angleee}
\end{eqnarray}

From these equations, one sees that a successful 3+1 model 
has two requirements: 1) that individual appearance and
disappearance data sets be well described by Eqs.~\ref{osceq} and
\ref{disosceq}, and 2) that all of the data sets together map onto the
same three  parameters:  $U_{e 4}$,  $U_{\mu 4}$, and
$\Delta m^2_{41}$.   We will show in Sec.~\ref{threeplusone} that the LSND
and MiniBooNE results fit poorly within a global 3+1 analysis.

This motivates extension to a 3+2 model, with two sterile
states and seven oscillation parameters.     In addition to the three
parameters from 3+1, 
the 3+2 model introduces
 an extra mass splitting, $\Delta m^2_{51}$, 
 the mixing elements, $|U_{e5}|$ and  $|U_{\mu 5}|$, 
and a complex phase, $\phi_{54}$.   The complex phase 
allows for significant $CP$ violation, manifested as a difference
between the neutrino and antineutrino appearance oscillation probabilities.    
However, because this is an interference effect, $CP$ violation will
only be significant if
$\Delta m^2_{51} /\Delta m^2_{41}\lesssim \mathcal{O}(10)$.
$CP$ violation will allow the signal for neutrino oscillations to be softer than for
antineutrino oscillations, which we will show is preferred by the
MiniBooNE data in Sec.~\ref{threeplustwo}.

Three sterile neutrino states, one for each active flavor generation,
seems most natural.  However, a 3+3 model is  
not found to significantly improve on the 3+2 fits, and the conclusions
are very similar.  
Therefore, for lack of space, we do not include a discussion of 3+3
phenomenology or fits.   Interested readers should see Ref.~\cite{Ignarra}.

\section{Comparing The LSND and MiniBooNE Experiments \label{expts}}

Although both LSND and MiniBooNE had $L/E \sim 1$  m/MeV to probe
$\Delta m^2 \sim 1$~eV$^2$, the designs were quite different.  
MiniBooNE used a beam which was an order of magnitude higher in 
energy than that of LSND, leading to different event signatures, backgrounds, 
and systematic uncertainties.   Correspondingly, MiniBooNE was located
a distance that was approximately an order of magnitude farther from
the neutrino source.
For a brief comparison, see  Table~\ref{tab:compare}. 

\begin{table}
\caption{Overview of the LSND and MiniBooNE experiments.}
{\begin{tabular}{@{}ccc@{}} \toprule
Property&LSND&MiniBooNE \\
\colrule
Proton Energy&798 MeV&8000 MeV \\
Proton Intensity&1000 $\mu$A&4 $\mu$A \\
Proton Beam Power&798 kW&32 kW \\
Protons on Target&28,896 C&284 C \\
Duty Factor&$6 \times 10^{-2}$&$8 \times 10^{-6}$ \\
Total Mass&167 tons&806 tons \\
Neutrino Distance& 29.8 m&541 m \\
Events for 100\% $\nu_\mu \rightarrow \nu_e$ Transmutation&33,300&
128,077 \\
\botrule
\end{tabular}
}
\label{tab:compare}
\end{table}

\subsection{LSND \label{lsndexpt}}

\begin{figure}
\centerline{\includegraphics[width=3.in,angle=0]{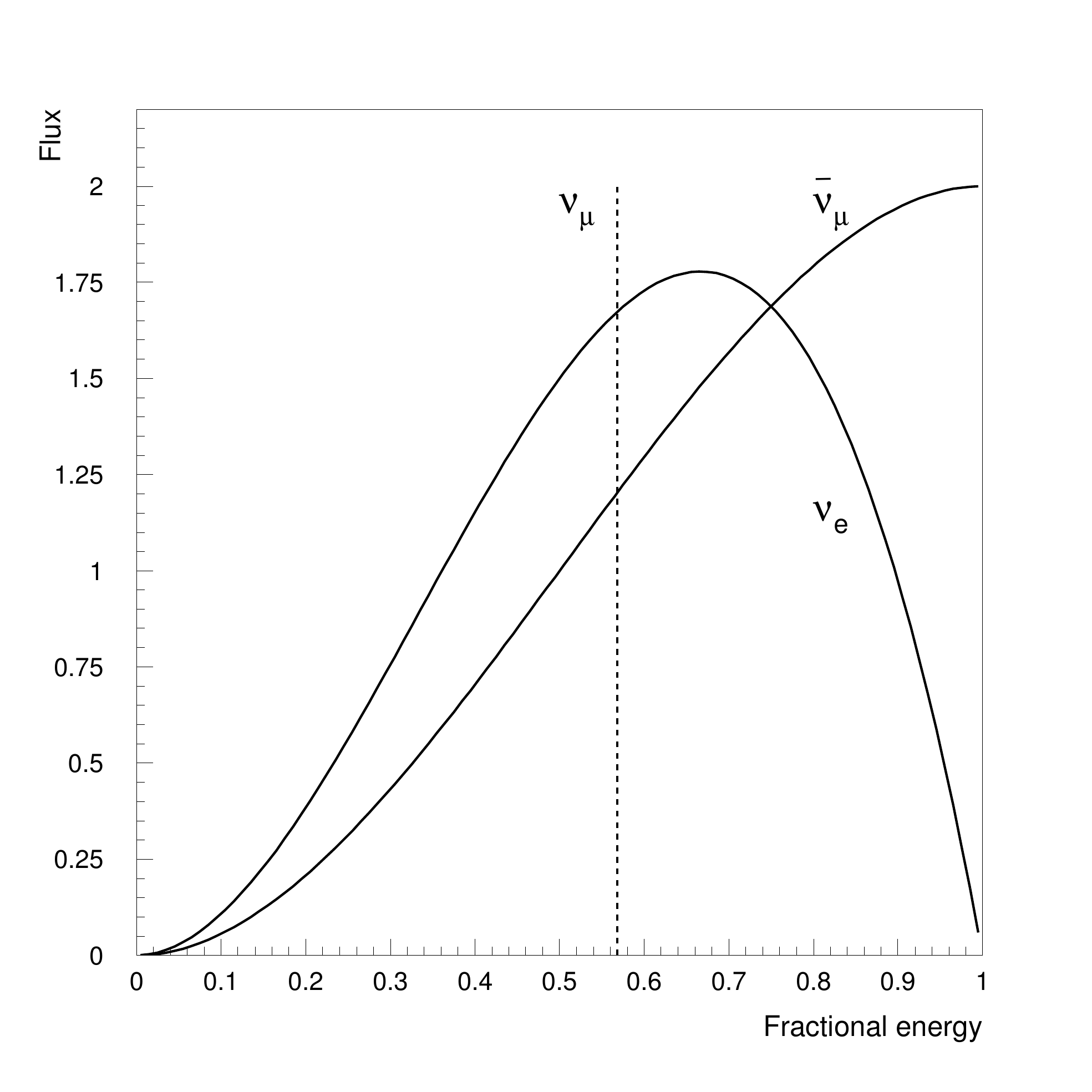}}
\caption{
The neutrino
energy spectra from $\pi^+$ and $\mu^+$ DAR.
}
\label{figure1}
\end{figure}

The LSND beam was produced at the Los Alamos National Laboratory
LAMPF/LANSCE accelerator.  A  
1 mA proton beam at 798 MeV impinged on 
the target/dump system with duty factor of $6 \times 10^{-2}$
\cite{lsndPRD} to produce 
pions and muons that decayed at rest.  The  decay-at-rest (DAR) neutrino flux,
shown in  Fig.~\ref{figure1}, arises from
stopped $\pi^+ \rightarrow \nu_\mu
\mu^+$ decay followed by stopped $\mu^+ \rightarrow \bar \nu_\mu \nu_e
e^+$.    The sister decay chain from stopped $\pi^-$ is highly suppressed through pion capture on target
nuclei.   The $\bar \nu_e$ intrinsic background at LSND was only $\sim 8 \times 10^{-4}$
of the $\bar \nu_{\mu}$ flux
in the energy range of the analysis \cite{bib:burman}. 

The $\bar \nu_e$ in the beam, either due to signal or background, could interact via ``inverse
beta decay''  (IBD),
$\bar \nu_e p \rightarrow e^+ n$, in the mineral oil target of the
LSND detector.   This reaction has a twofold signature of a prompt 
positron  and a correlated 2.2 MeV $\gamma$ from neutron capture.
Although the target oil was lightly doped with scintillator, the Cherenkov
ring could still be reconstructed, allowing the determination of the
energy and angle of the outgoing
positron.  The $\nu_\mu$ and $\bar \nu_\mu$ energies 
were below threshold
for charged-current (CC) muon production; thus, 
only neutral-current (NC) events were produced.
The $\nu_e$ in the beam produced
$\nu_e +^{12}{\rm C} \rightarrow ^{12}{\rm
  N}_{gs} + e^-$ events, 
where {\it gs} indicates ground state. This was not confused with the IBD signal, as there was no correlated
neutron capture in these events.  The $\nu_e$ events were used in a disappearance
study discussed in Sec.~\ref{disresults}.

\begin{figure}
\centerline{\includegraphics[height=3.in,angle=0]{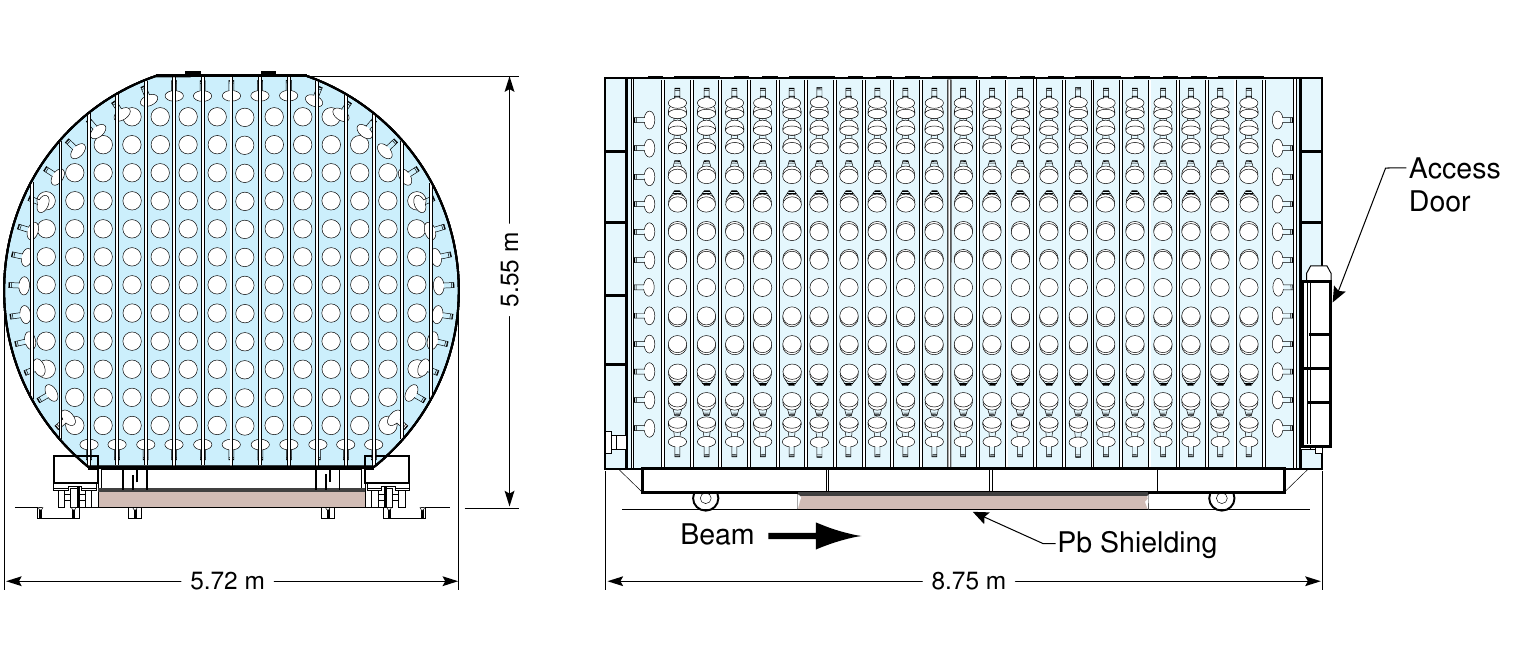}}
\caption{
A schematic drawing of the LSND detector.
}
\label{figure2}
\end{figure}

Fig. \ref{figure2} illustrates the LSND detector \cite{lsndNIM}, which 
consisted of a cylindrical tank,
8.3 m long by 5.7 m in diameter with center located 29.8 m from the neutrino source.
1220 8-inch Hamamatsu photomultiplier tubes (PMTs)
covered
25\% of the tank surface area.
The tank was filled with 167 tons of liquid scintillator consisting of mineral
oil and 0.031 g/l of b-PBD.
A typical 45-MeV electron from a CC interaction produced 
$\sim 1500$ photoelectrons, of which $\sim 280$ photoelectrons were in the
Cherenkov cone.
PMT time and pulse-height signals were used to reconstruct the
track with an average RMS position resolution of $\sim 14$ cm,
an angular resolution of $\sim 12^\circ$, and an energy resolution of
$\sim 7\%$ at the Michel electron endpoint of 52.8 MeV.

The veto shield, consisting of a 15-cm layer of liquid
scintillator in an external tank and 15 cm of lead shot in an internal
tank, enclosed the detector on all sides except the bottom \cite{veto}.
The active veto
tagged cosmic-ray muons that stopped in the lead shot.
Additional counters were placed below the tank
after the 1993 run to reduce
cosmic-ray background entering through the bottom support structure.
A veto inefficiency $<10^{-5}$ was achieved
for incident charged particles, and the veto introduced a  $0.76 \pm 0.02$
deadtime.

\subsection{MiniBooNE \label{mbexpt}}

\begin{figure}
\centering
\includegraphics[width=10.5cm,
scale=1.0]{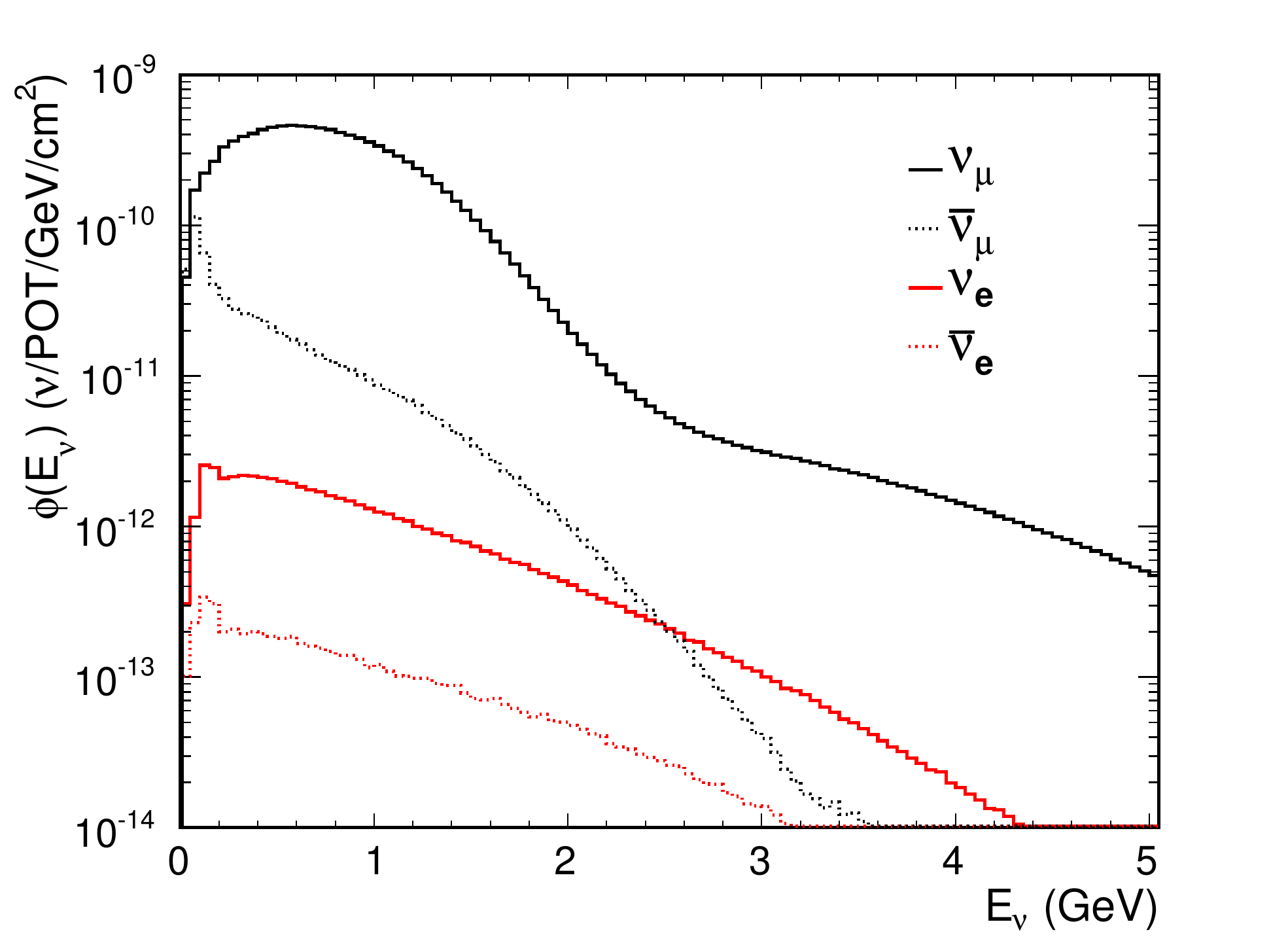}
\hspace{1cm}
\includegraphics[width=10.5cm,
scale=1.0]{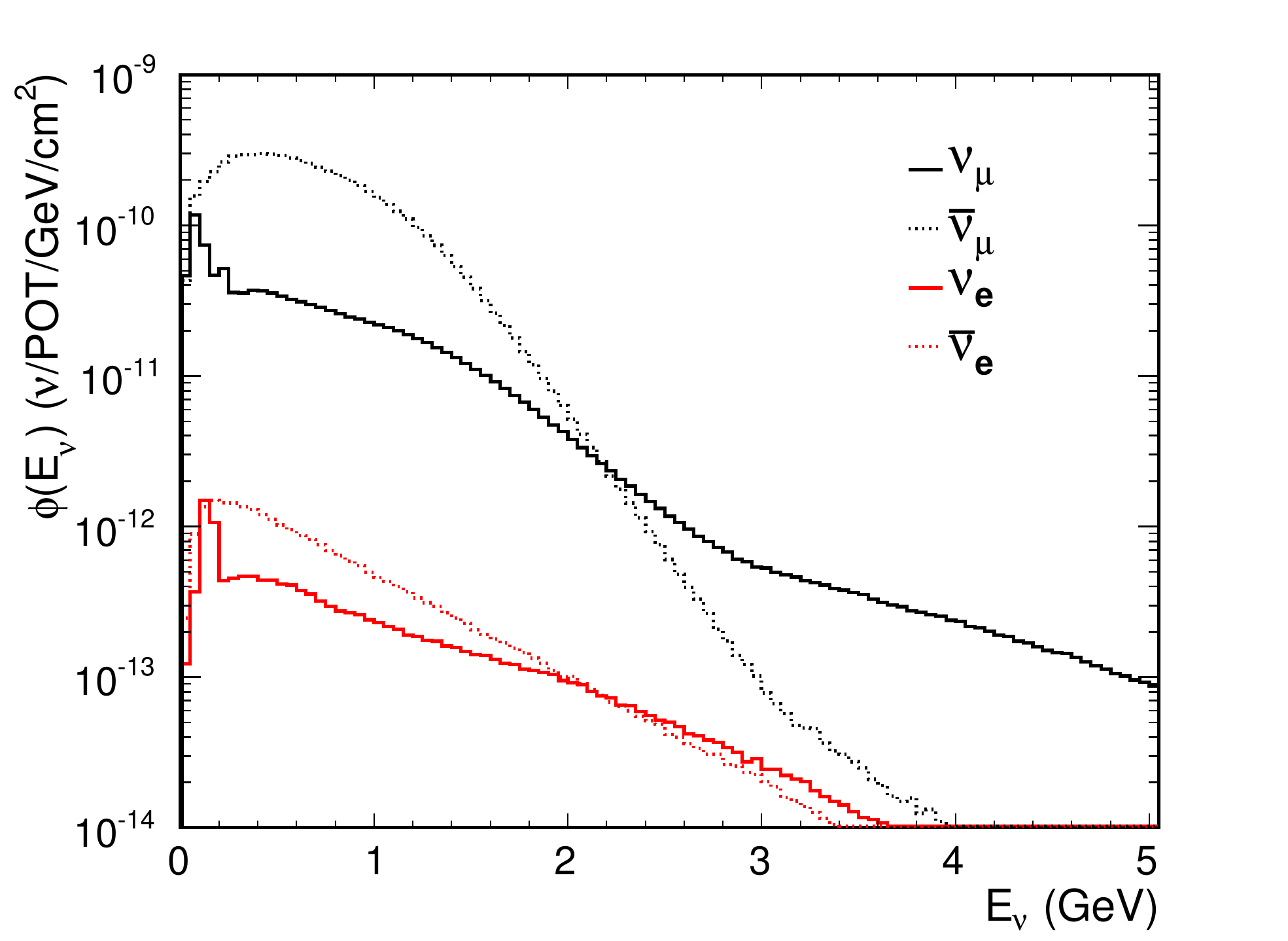}
\parbox{7in}{
\caption{\em The estimated neutrino fluxes for neutrino mode (top plot) and
antineutrino mode (bottom plot).} 
\label{figure3}}
\end{figure}

In contrast to LSND, MiniBooNE made use of a ``conventional neutrino 
beam'' at Fermilab where protons hit a target and produce mesons 
that decay in flight.  The Booster accelerator fed 8-GeV kinetic energy
protons to a 71-cm long Be target located at the upstream end of a magnetic
focusing horn. The horn pulsed with a current of 174 kA.  Depending on
the polarity, this either focused $\pi^+$ and $K^+$ or 
$\pi^-$ and $K^-$ into a 50-m decay pipe. 
The neutrinos or antineutrinos produced from the meson decays 
could interact in the MiniBooNE
detector, located 541 m downstream of the Be target. 

Fig. \ref{figure3} shows the neutrino fluxes for this Booster Neutrino
Beam (BNB) used by MiniBooNE, for neutrino
mode and antineutrino mode running \cite{mb_flux}. The fluxes are
fairly similar for the two modes, both
in energy and in the intrinsic electron-neutrino background, which is $\sim$0.6\%. 
However, as with all conventional neutrino
beams,  the wrong-sign contribution to the
flux in antineutrino mode ($\sim 18\%$) is much larger than in neutrino mode
($\sim 6\%$). 

\begin{figure}
\centerline{\includegraphics[height=3.5in]{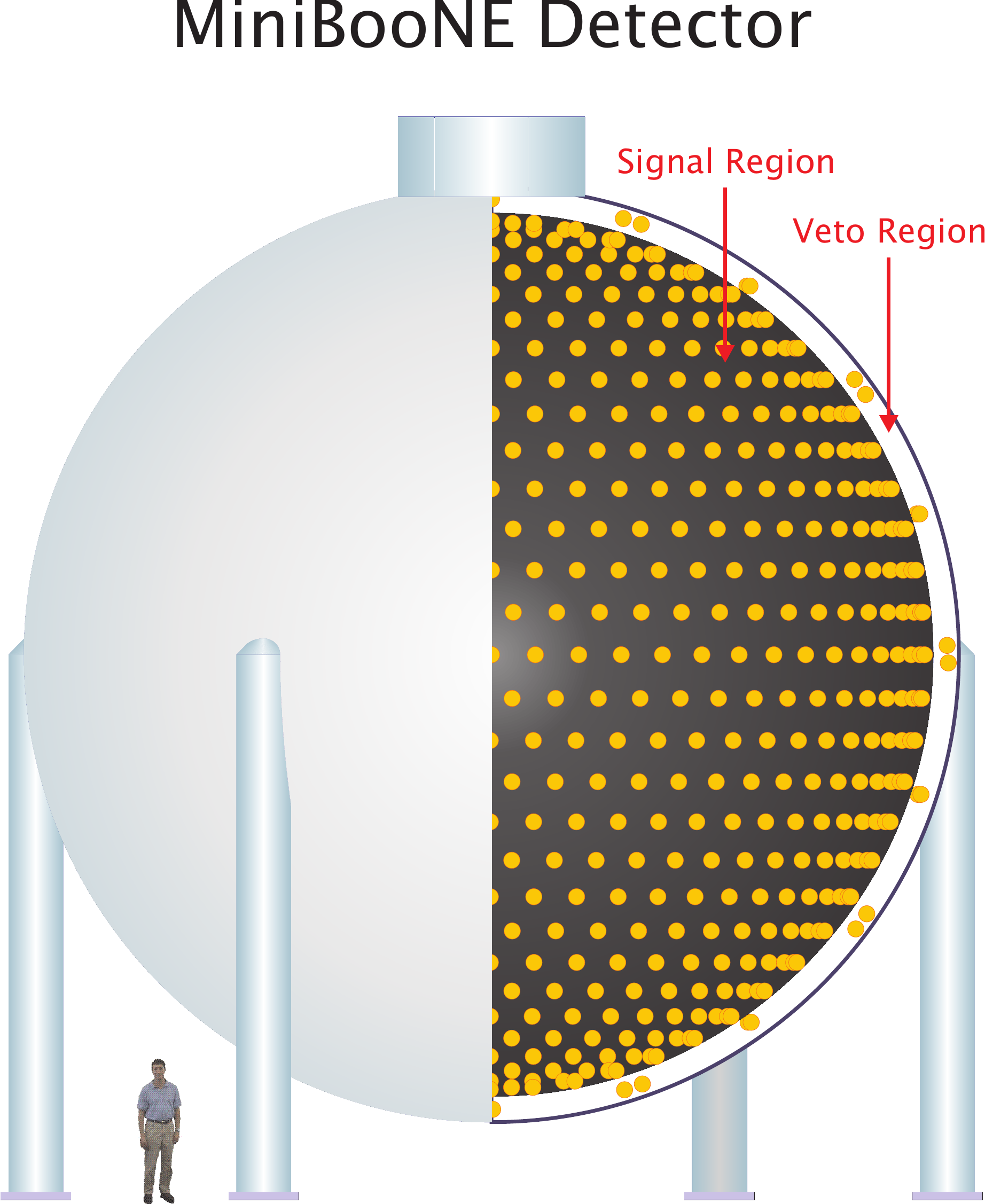}}
\caption{\label{figure4} \em A schematic drawing of the MiniBooNE detector.}
\end{figure}

MiniBooNE \cite{mb_detector} utilized a 12.2-m diameter
spherical tank filled with 806 tons of mineral oil ($CH_2$) with no
scintillator doping, shown schematically in Fig. \ref{figure4}.
A total of 1280 8-inch PMTs covered 10\% of the surface
area of the target region, which was painted black to reduce
reflections.  
The fiducial volume within the target had a 5-m radius and
corresponded to 450 tons.
The target region was optically isolated from a
full-coverage veto that was 30 cm thick and  
contains 240 veto PMTs.    White reflective paint on the veto
walls led to high reflection and excellent efficiency for cosmic-muon detection of $>99.9$\%.

At BNB energies, the important contributions to the neutrino and
antineutrino CC cross sections are quasi-elastic (QE) scattering and single 
pion production, typically from $\Delta$ baryon production.
A review of 
these interactions is provided in Ref.~\cite{SamAnnualReview}, and we
note that MiniBooNE has provided a great deal of new data related to
the cross-section measurements. 
The targets were nucleons in the carbon atoms and free protons
associated with the hydrogen atoms.
Signal identification and background rejection relied on the measured
characteristics of the observed Cherenkov rings.   As a result, the
analysis proceeded quite differently to that of LSND.
A key aspect
of MiniBooNE is that the backgrounds are very well understood and are
constrained directly from measurements in the detector as described below.

\section{Appearance Results from LSND and MiniBooNE \label{results}}

Both the LSND and MiniBooNE experiments have reported muon-flavor to
electron-flavor appearance signals.  This section will show
that the results of each search are individually consistent with the two-flavor
oscillation phenomenology introduced in Sec.~\ref{pheno}.  
However, in Sec.~\ref{sterilecon} we show that an extended model is
required to explain the combined data sets.

\subsection{LSND \label{lsndres}}

LSND presented a number of incremental results throughout the run
\cite{lsndPRL95,lsndPRL96,lsndPRL98}, and the final results were presented in a
comprehensive paper in 2001 \cite{lsndPRD}.    In this section, we mainly review
the primary oscillation analysis but briefly consider several cross-check analyses performed
to address the consistency of the result.

\subsubsection{LSND Oscillation Analysis \label{lsndprime}}
 
Due to the poor duty factor, the raw event sample of LSND had a high
cosmic-ray content, and so initial ``Reduction Criteria'' were applied.   The first step
was a prompt energy
requirement of $E_e>20$ MeV.   
Timing cuts on target and veto shield activity
further reduced the cosmic background.
Next, ``Electron Selection Criteria'' were applied.  These cuts
isolated candidate events in time,  required a reconstructed 
event vertex greater than 35 cm from the faces of the PMTs, 
and selected on particle ID parameters derived from the position and timing of PMT
hits as described in Ref.~\cite{lsndPRD}.
The analysis also required $E_e<60$ MeV to isolate the DAR
sample from decay-in-flight events.
From the tagged Michel electron sample from cosmic-muon decay,
the efficiency for the Electron Selection was $0.42
\pm 0.03$.    

Next, the coincidence with a 
2.2 MeV $\gamma$ from neutron capture was required. 
The task was to distinguish true neutron captures
from accidental $\gamma$s from radioactivity. To this
end,  LSND introduced the 
ratio, $R_\gamma$, of the likelihood that the $\gamma$
is correlated divided by the likelihood that the $\gamma$ is
accidental, which depended upon three quantities: the number of hit
PMTs, since the multiplicity is proportional to the
$\gamma$ energy; the distance between the reconstructed
$\gamma$ position
and positron-candidate position; and the time interval between the $\gamma$ and
positron candidate.   

The $R_\gamma$ distribution of the events passing Electron Selection 
was fit to templates of the correlated signal and accidental
backgrounds with floating normalization, yielding a $\chi^2$/dof =
10.7/9.    From this, $117.9\pm 22.4$
$\bar \nu_e$ events were found to be in the sample.
Of these,  $19.5\pm 3.9$ and $10.5 \pm
4.6$ are predicted to be from intrinsic $\bar \nu$ sources from
$\mu^-$ decay at rest and $\pi^-$ decay in flight, respectively \cite{numubkgd}.   
Thus, the LSND signal excess corresponds to $87.9 \pm 22.4 \pm 6.0$ events.
For comparison, from the expected candidate rate with 100\% transmutation of the
$\bar\nu_\mu$ flux, one expects $33,300 \pm 3,300$ events.
Interpreting the excess as oscillations in a two-neutrino model, the
probability is $(0.264 \pm 0.067 \pm 0.045)\%$. 
                                                                                
\begin{figure}
\centerline{\includegraphics[height=3.in,angle=0]{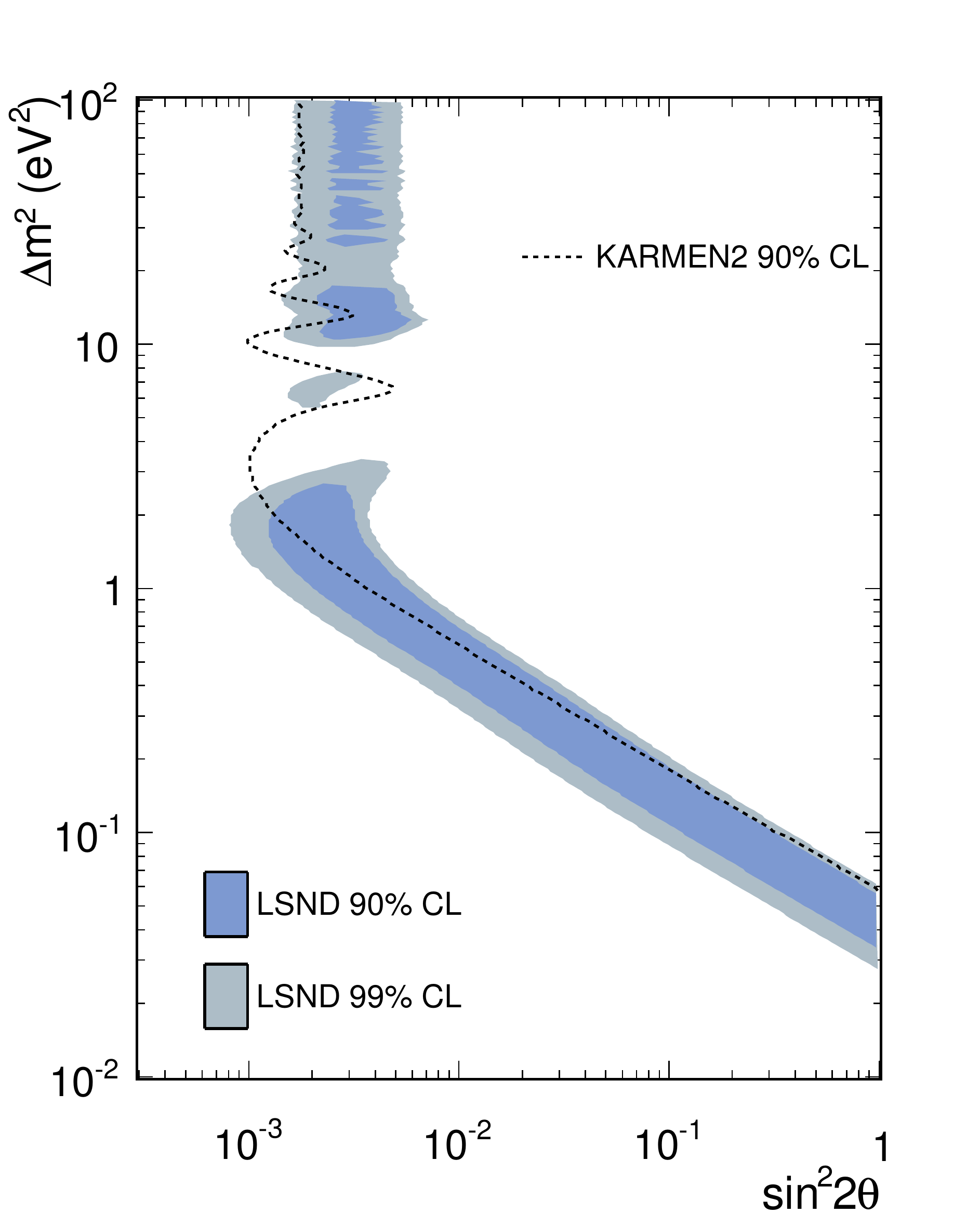}}
\caption{
The $(\sin^22\theta,\Delta m^2)$
oscillation parameter fit
for the entire LSND data set.
The inner
(outer) regions correspond to 90\% (99\%) CL allowed regions.
90\% CL limits from the KARMEN2\cite{Karmen2}
experiments are also shown.}
\label{figure5}
\end{figure}

Using Eq.~\ref{osceq}, a fit is performed 
for $\bar \nu_\mu \rightarrow \bar \nu_e$ appearance by calculating 
the likelihood ($\mathcal{L}$) in the $(\sin^22\theta,\Delta m^2)$ plane to
extract the favored oscillation parameters.  The 
three-dimensional contour in $(\sin^22\theta,\Delta m^2,  \mathcal{L})$  is sliced to find
the LSND allowed oscillation region.
The result is shown in Fig. \ref{figure5}, where the  inner 
(outer) region corresponds to a 90\% (99\%) CL.  

In the same timeframe as the LSND run, the KARMEN experiment \cite{Kardet}
took data with a DAR beam at the ISIS facility at the Rutherford
Laboratory.  A key difference with respect to LSND is the
KARMEN location at 17.7 m from the target at a 100$^\circ$ angle to the proton beam. 
KARMEN did not observe an oscillation signal \cite{Karmen2} and obtained
the 90\% CL limit shown in Fig.~\ref{figure5}.  KARMEN restricts part of the
LSND region and, through a joint analysis with LSND, was used
to determine a combined allowed region for the two experiments \cite{ref:joint}.

The most controversial cuts in the DAR analysis 
have been those on the fiducial volume.  Questions arose
because of an apparent up-down asymmetry in the 
first LSND result, presented in 1995.   The 
result, which also used a stricter energy cut, $E_e>36$
MeV, than the final analysis,  had only nine
candidate events, with six at $Y<0$ and three at $Y>0$ \cite{lsndPRL95}. 
Although this is not a highly improbable $Y$ distribution,
concern was raised 
because the top of the detector had
complete veto coverage, while the bottom did not.  Continued
running smoothed the statistical fluctuation.   Table
\ref{exercises} provides the oscillation probabilities for the final
event sample with exercises in varying the
fiducial cuts, showing that the signal is resilient to these cuts.   

\begin{table}[t]
\caption{Exercises in restricting the prompt events \cite{lsndPRD}.  The analysis 
used a right handed coordinate system with beam along $Z$ and $Y$
along the vertical axis.  $D$ is distance from the PMT face.}
{\begin{tabular}{@{}cc@{}} 
Selection & Oscillation Probability in \% \\ \toprule
Primary Analysis & $0.264\pm 0.067 \pm 0.045$ \\
Primary + $D>50$ cm and $Y>-50$ cm & $0.252\pm0.071\pm0.045$ \\
Primary + $D>75$ cm &  $0.193\pm0.055\pm0.045$ \\
Primary + $Y>-120$ cm & $0.293\pm0.069\pm0.045$\\
\botrule
\end{tabular}
}
\label{exercises}
\end{table}

\begin{figure}
\centerline{\includegraphics[height=3.in,angle=0]{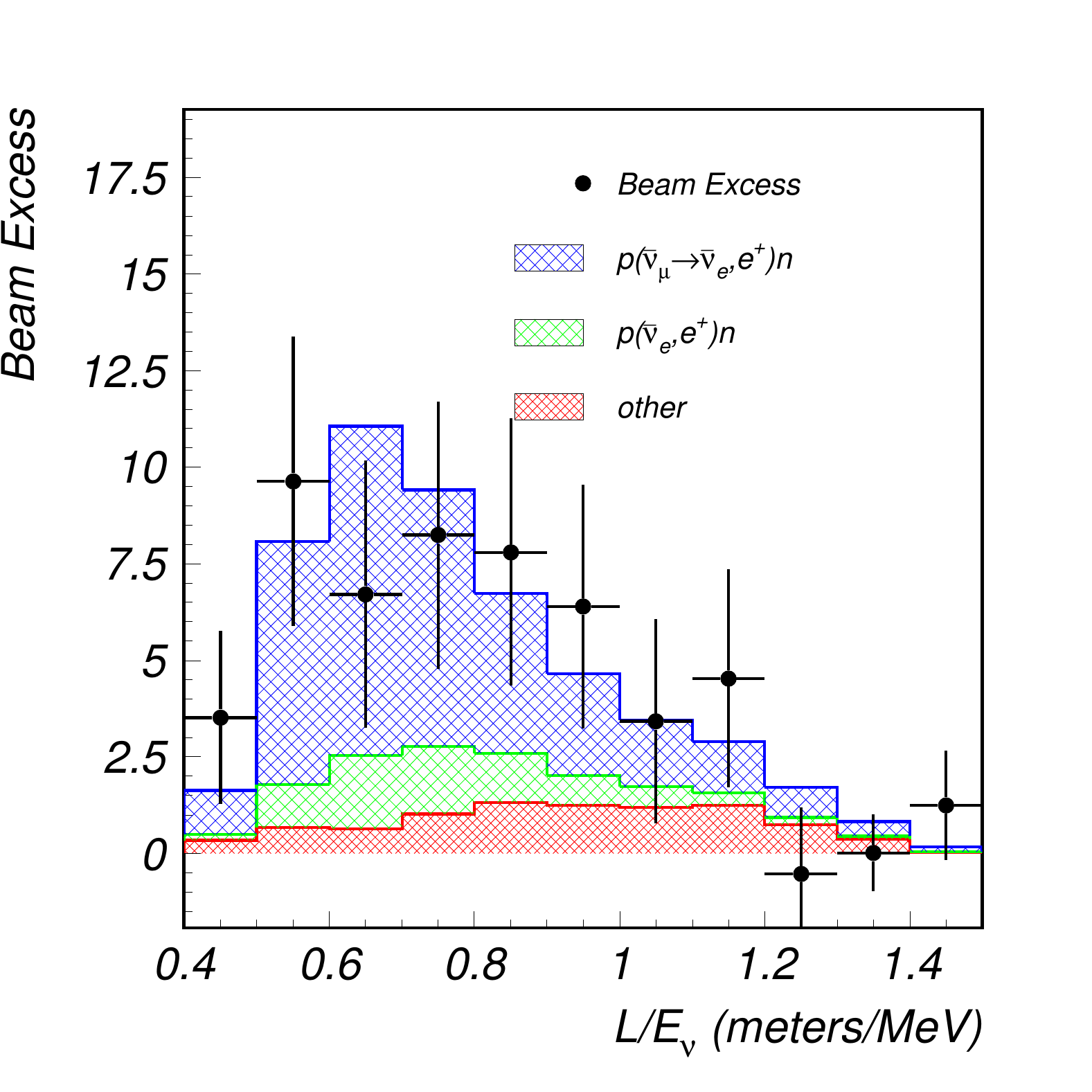}}
\caption{
The LSND $L/E$ distribution
for events with $R_\gamma >10$ passing the Electron Selection criteria.
}
\label{figure6}
\end{figure}

Another useful cross-check maintains the Electron Selection cuts but employs 
an $R_\gamma>10$ cut rather than the template
fit.    This isolates a very clean signal, revealing 
the hallmark $L/E$ distribution evident in Fig.~\ref{figure6}.  
The event excess is $32.2\pm9.4\pm2.3$ and 
the probability that this is a statistical fluctuation is
$1.1\times 10^{-4}$.

\subsection{MiniBooNE \label{MBres}}

The MiniBooNE experiment ran for ten years, from 2002 until 2012, switching between neutrino
and antineutrino mode running.  The final data sample corresponds to $6.46 \times 10^{20}$
($11.27 \times 10^{20}$)
protons on target (POT) in neutrino (antineutrino) mode.  
MiniBooNE searched for $\nu_\mu \rightarrow \nu_e$ 
(or $\bar \nu_\mu \rightarrow \bar\nu_e$) oscillations by
measuring the rate of $\nu_e n \rightarrow e^- p$ 
(or $\bar\nu_e p \rightarrow e^+ n$) charged-current quasi-elastic
(CCQE) events and
testing whether the measured rate was consistent with the estimated
background rate.   For these events, the incoming $\nu/\bar\nu$ energy is 
approximated according to the QE formula:
\begin{equation}
E_{\nu_{mode}/\bar\nu_{mode}}^{QE}=\frac{2(M_{n/p}^{\prime})E_e-((M_{n/p}^{\prime})^2+m_e^2-M_{p/n}^2)}
{2\/[(M_{n/p}^{\prime})-E_e+\sqrt{E_e^2-m_e^2}\cos\theta_e]},\label{eq:recEnu}
\end{equation}
where $M_n$, $M_p$, and $m_e$ are the neutron, proton, and electron masses,
and $E_e$ and $\cos \theta_e$ are the energy and angle of the
outgoing electron, respectively.  The adjusted neutron/proton mass is defined 
as $M_{n/p}^{\prime}=M_{n/p}-E_B$,
with binding energy $E_B=34\pm9$~MeV.

A number of papers were published over
this period documenting the oscillation analyses with various data samples and stages of
analysis.  The first oscillation publication was in 2007 \cite{MBnu} on the neutrino mode search,
and it showed that there was no excess of events
with $E_\nu^{QE} > 475$~MeV, which was somewhat inconsistent with the LSND result for antineutrinos.
This was followed by a paper in 2009 \cite{unexplained} showing that there was an 
unexplained excess of $\nu_e$ events for MiniBooNE neutrino running
below $475$~MeV, prompting speculations of $CP$ violation such as that
included in 3+2 models.  At this point, MiniBooNE switched to mainly antineutrino mode
running, with early results presented in 2009 \cite{MBnubar09} and 2010 \cite{MBnubar10}.  
The antineutrino oscillation results for the
full data sample were recently posted \cite{MBnubar} and show an antineutrino excess consistent with
the LSND signal region.
  
\subsubsection{MiniBooNE Oscillation Analysis \label{mbevsel}}

The MiniBooNE oscillation search can be broken into two main 
components.  The first component uses analysis cuts to isolate a fairly pure sample 
of electron neutrino events, and the second component uses
background estimates and measurements to determine 
the size and uncertainty of a possible excess from oscillations.
The main backgrounds to a $\nu_e$ oscillation signal can be divided
into two types: 1) single $\gamma$ events that mimic an outgoing electron
and 2) intrinsic $\nu_e$-induced events that are identical to the oscillation signal.
The single $\gamma$ backgrounds are important since the MiniBooNE detector
cannot separate events with a single $\gamma$ from those with an outgoing 
electron.
These backgrounds include NC $\pi^0$ events
where one of the decay $\gamma$ rays is unobserved,  radiative $\Delta \rightarrow
N \gamma$ events, and $\gamma$'s from external neutrino interactions outside of the
detector.  The intrinsic  $\nu_e$ backgrounds are from the decays of secondary muons
plus charged and neutral kaons produced in the
primary production target and shielding.

To select candidate $\nu_e$ CCQE events, an initial selection is
first applied:
$>200$ tank hits, $<6$ veto hits, reconstructed time within the
neutrino beam spill, reconstructed vertex radius $<500$ cm,
and visible energy $E_{vis}>140$ MeV.  With these cuts, the
cosmic-ray backgrounds are negligible.  
It is then required
that the event vertex be reconstructed assuming an outgoing electron and the track
endpoint reconstructed assuming an outgoing muon
occur at radii $<500$ cm and $<488$ cm, respectively, to
ensure good event reconstruction and efficiency for possible muon decay
electrons.  One remaining background from neutrino interactions in the material
surrounding the detector is substantially reduced using correlated energy and
topology cuts, and the subsequent rate is measured from isolated background 
events that have low energy, large radius, and a topology that points into the detector.

After the selection cuts, the surviving events are reconstructed under four hypotheses: 
a single electron-like Cherenkov ring, a single muon-like ring, 
two photon-like rings with unconstrained kinematics, and two photon-like rings consistent with the decay of a 
$\pi^0$. The assessment of detector
response to these hypotheses uses a detailed model of extended-track light production 
and propagation in the tank to predict the charge and time of hits on each PMT. 
This reconstruction yields a position, direction, and energy resolution for $\nu_e$
events of 22 cm, 2.8$^\circ$, and 11\%, respectively, and a $\pi^0$ mass resolution of
20 MeV/c$^2$.

Particle identification (PID) cuts are then applied
to reject muon and $\pi^0$ events.  The PID uses energy-dependent cuts on the likelihood
ratios for the four above hypotheses, specifically $\log(L_e/L_\mu)$, $\log(L_e/L_{\pi^0})$,
and $M_{\gamma\gamma}$.  These PID cuts substantially reduce the $\gamma$ backgrounds but
have a high efficiency ($55 \pm 3$\%) for $\nu_e$-induced events.

All of the MiniBooNE backgrounds are constrained by {\it in-situ} measurements.  The $\nu_\mu$
inclusive CC background is verified by comparing the Monte Carlo (MC) prediction to the large sample 
of tagged events with a Michel decay electron.  Over 99\% of the NC $\pi^0$ events are correctly
reconstructed as two $\gamma$ from $\pi^0$ decay and can be used to constrain the 
background where one $\gamma$ is missed.  This sample can also be used to constrain the 
radiative $\Delta \rightarrow N \gamma$ background.  The intrinsic $\nu_e$ background events from muon
decay are directly related to the observed $\nu_\mu$ events since both come from a common
$\pi^\pm$ decay chain.  MiniBooNE uses a combined fit 
of the observed $\nu_\mu$ and $\nu_e$ events, including correlations, 
to effect this constraint.  The other
major source of background $\nu_e$ events is K$^+$ decay, where the K$^+$ rate has been measured 
using the high-energy events in the SciBooNE detector located near the end of the BNB decay pipe \cite{ChengKProd}.

Table \ref{signal_bkgd} shows the
expected number of candidate $\nu_e$ and $\bar \nu_e$
CCQE background events with
$E_\nu^{QE}$ between $200 - 1250$ MeV
after the complete event selection of the final analysis.
After applying the above mentioned $\nu_\mu$ constraint, 
the total expected backgrounds for neutrino mode and antineutrino mode
are $790.0 \pm 28.1 \pm 38.7$ and $399.6 \pm 20.0 \pm 20.3$ events, 
respectively, where the first (second) error is statistical (systematic).
The number of data events after all cuts is 952 for neutrino mode and 478 for
antineutrino mode, giving data excesses of $162 \pm 47.8$ and $78.4 \pm 28.5$ events
for the two modes, where the error includes both statistical and systematics
uncertainties.

\begin{table}[htpb]
\caption{\label{signal_bkgd} \em The MiniBooNE
expected (unconstrained) number of events
for the $200<E_\nu^{QE}<1250$~MeV neutrino oscillation
energy range from all of the backgrounds in the $\nu_e$ and $\bar{\nu}_e$
appearance analysis and for the LSND expectation of 0.26\% oscillation 
probability averaged over neutrino energy
for both neutrino mode and antineutrino mode. Also shown are the total number of data events and the total constrained background (From Ref.~\cite{MBnubar}.)
}
\vspace{0.2in}
\small
\begin{tabular}{c|cc|cc}
\hline
Process&Neutrino Mode&  &Antineutrino Mode&   \\
&200-475 MeV&475-1250 MeV&200-475 MeV&475-1250 MeV \\
\hline
$\nu_\mu$ \& $\bar \nu_\mu$ CCQE &25.4&11.7&8.8&4.1  \\
NC $\pi^0$ &181.2&71.1&85.4&26.9 \\
NC $\Delta \rightarrow N \gamma$ &66.9&19.9&26.4&8.3 \\
External Events &23.9&11.5&10.8&4.5 \\
Other $\nu_\mu$ \& $\bar \nu_\mu$ &28.8&16.4&13.8&8.5 \\
\hline
$\nu_e$ \& $\bar \nu_e$ from $\mu^{\pm}$ Decay &58.7&155.3&27.2&64.2 \\
$\nu_e$ \& $\bar \nu_e$ from $K^{\pm}$ Decay &17.2&79.5&15.5&35.7 \\
$\nu_e$ \& $\bar \nu_e$ from $K^0_L$ Decay &6.3&21.1&10.1&41.4 \\
Other $\nu_e$ \& $\bar \nu_e$ &0.8&2.2&2.5&4.1 \\
\hline
Total Unconstrained Background &409.1&388.7&200.5&197.7 \\
\hline
0.26\% $\bar{\nu}_{\mu}\rightarrow\bar{\nu}_e$ &50.4&182.7&23.7&76.3 \\
Total Constrained Background&401.3&388.8&203.3&196.3 \\
Number of Data Events&544&408&257&221 \\
\hline
\end{tabular}
\normalsize
\vspace{0.1in}
\end{table}

\begin{figure}[tbp]
\vspace{-0.1in}\centerline{\includegraphics[angle=0, width=10.5cm]{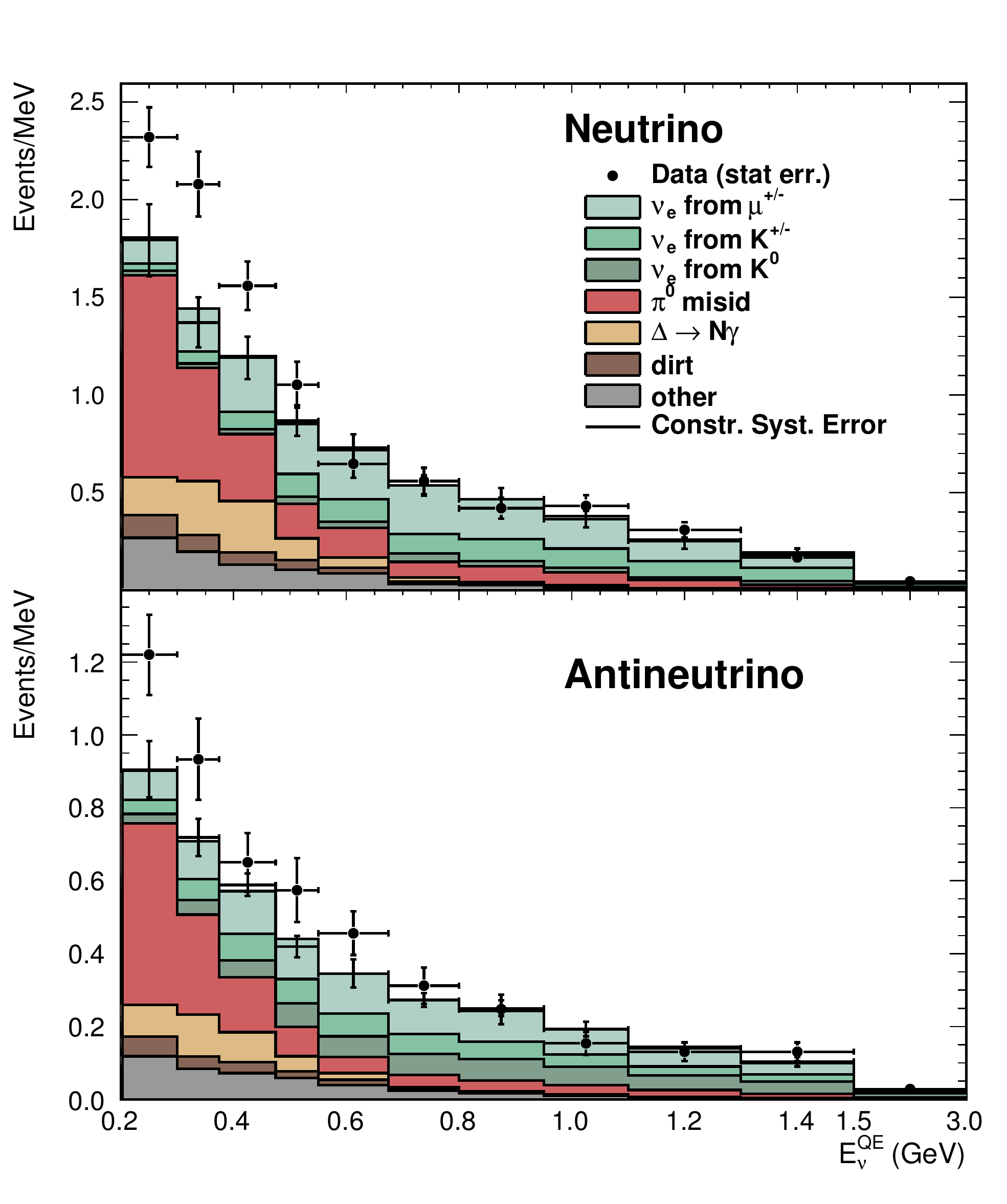}}
\vspace{0.1in}
\caption{The neutrino mode (top) and antineutrino mode (bottom) $E_\nu^{QE}$ distributions
for ${\nu}_e$ CCQE data (points with statistical errors) and background 
(histogram with systematic errors).}
\label{figure7}
\vspace{-0.2in}
\end{figure}

\begin{figure}[tbp]
\vspace{-0.1in}\centerline{\includegraphics[angle=0, width=10.5cm]{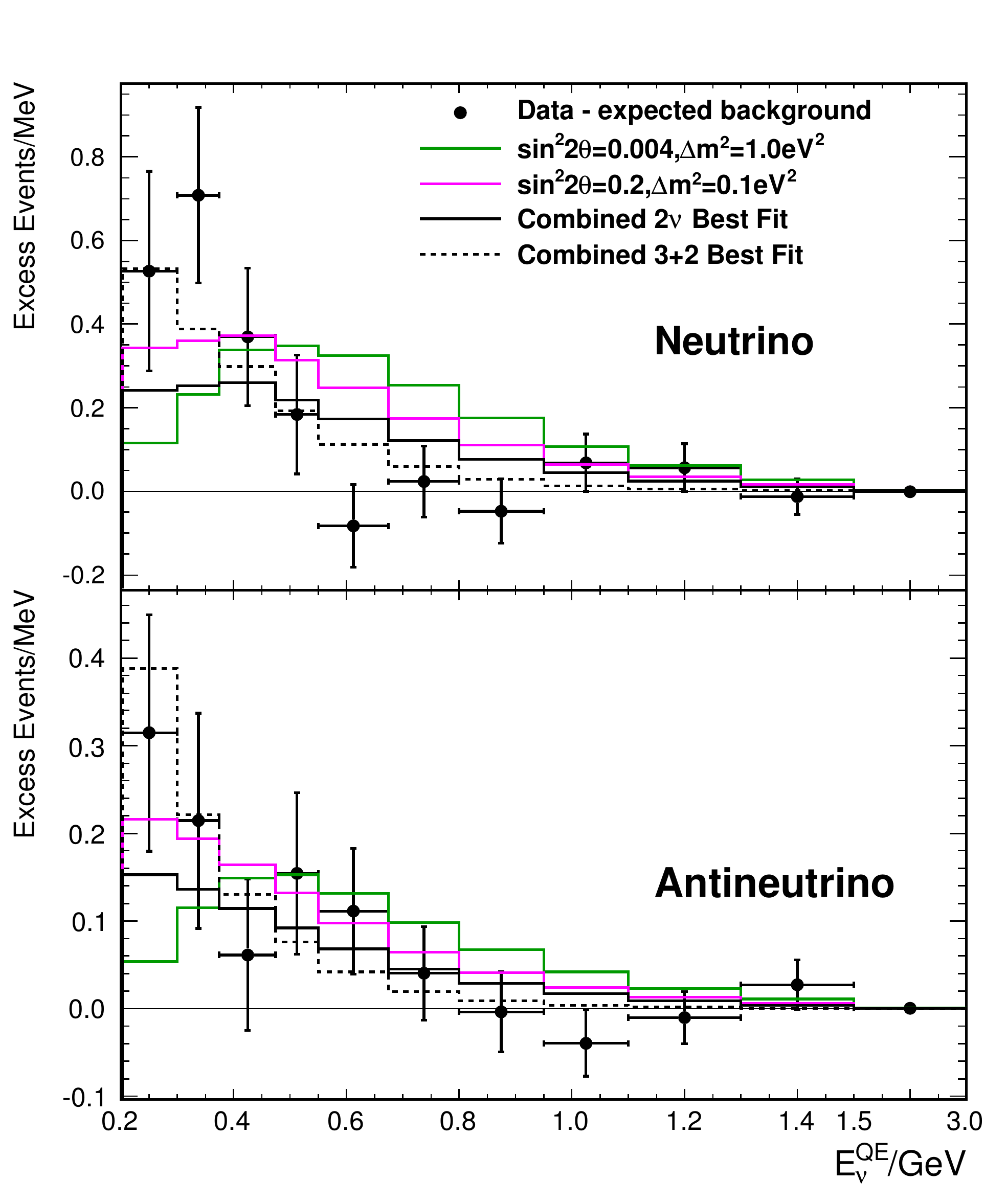}}
\vspace{0.1in}
\caption{The neutrino mode (top) and antineutrino mode (bottom)
event excesses as a function of $E_\nu^{QE}$. Also shown are the
expectations from the best two-neutrino and 3+2 joint oscillation fits
with $200<E_\nu^{QE}<3000$~MeV and from two reference values in the LSND
allowed region.
All known systematic errors are included in the systematic error estimate.}
\label{figure8}
\vspace{-0.2in}
\end{figure}

Fig. \ref{figure7} shows the reconstructed neutrino and antineutrino 
energy distributions
for candidate $\nu_e$ and $\bar \nu_e$ 
data events (points with error bars) compared to
the MC simulation (histogram with systematic uncertainties) \cite{MBnubar}.
Fig. \ref{figure8} shows the event excess as a function of 
reconstructed neutrino energy. 
For the neutrino data, the magnitude of the excess is similar to that expected from
the LSND antineutrino oscillation signal, 
but the shape shows a decided difference, 
being larger below 400 MeV and much smaller above 500 MeV.
The lack of a significant excess above 475 MeV is the source of the original MiniBooNE claim
of incompatibility with LSND.
In contrast, the antineutrino excess shows a similar magnitude and shape with
respect to the LSND predictions and is fully consistent with the LSND signal. 

\begin{figure}[tbp]
\vspace{-0.4in}
\centerline{\includegraphics[width=10.5cm]{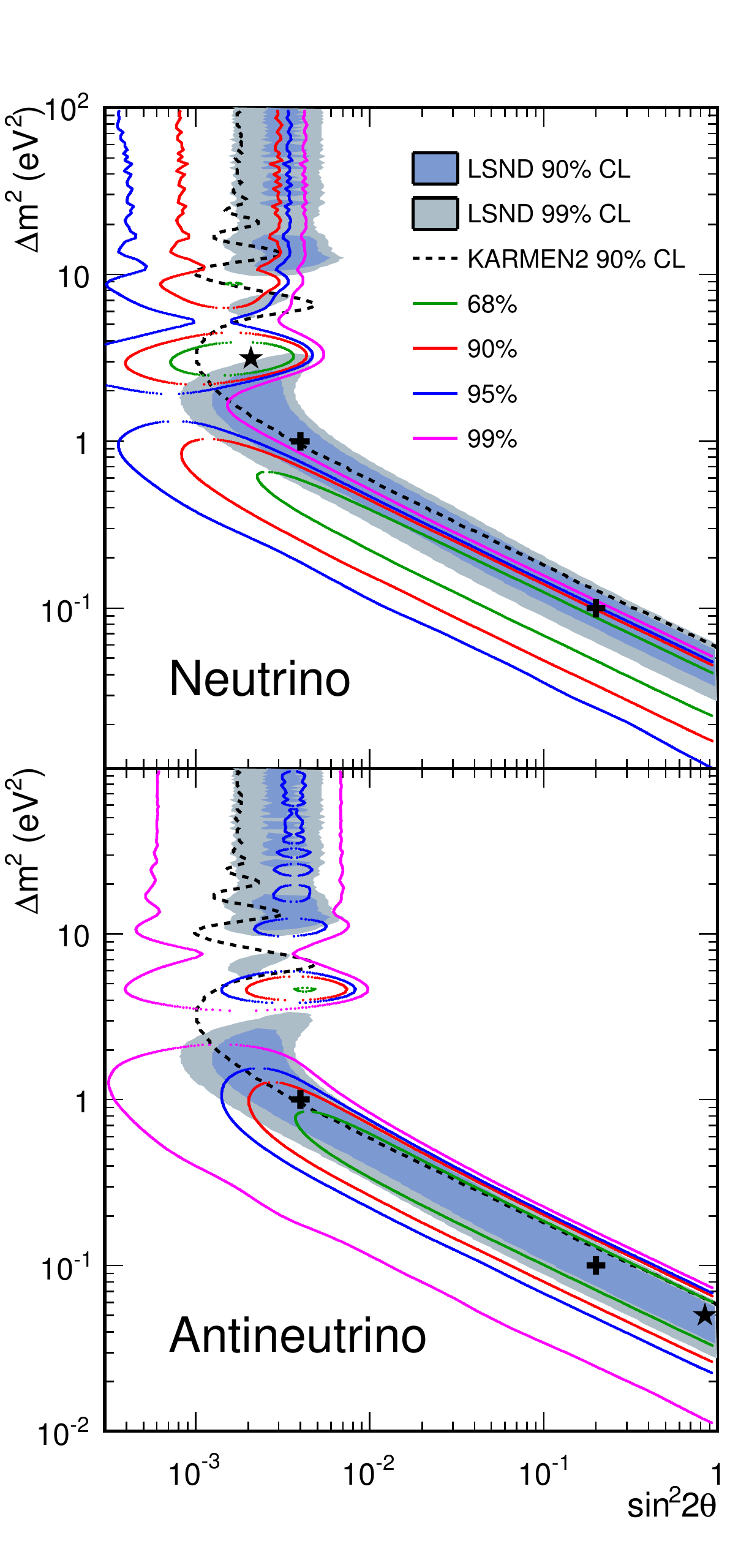}}
\vspace{-0.4in}
\caption{MiniBooNE allowed regions in combined neutrino and antineutrino mode for events with
$200<E^{QE}_{\nu}< 3000$~MeV within a two-neutrino ${\nu}_{\mu}\rightarrow{\nu}_e$ and
$\bar{\nu}_{\mu}\rightarrow\bar{\nu}_e$ oscillation model.
Also shown is the $\bar{\nu}_{\mu}\rightarrow\bar{\nu}_e$
limit from the KARMEN experiment \cite{Karmen2}.
The shaded areas show the 90\% and 99\% CL LSND
$\bar{\nu}_{\mu}\rightarrow\bar{\nu}_e$ allowed
regions. The black star shows the best fit point.}\label{figure9}
\vspace{-0.6in}
\end{figure}

The MiniBooNE neutrino and antineutrino data can be fit to a two-neutrino oscillation model, where the
probability is given by Eq. \ref{osceq}. 
The $\nu_e$ fit is constrained by the observed $\nu_\mu$ data by doing a combined fit of 
the observed $E_\nu^{QE}$ event distributions for muon-like and electron-like events.
The fit assumes the same oscillation probability for 
both the right-sign $\bar \nu_e$  and wrong-sign  $\nu_e$ and
no significant $\nu_\mu$, $\bar{\nu}_{\mu}$, $\nu_e$,
or $\bar \nu_e$ disappearance. 
Using a likelihood-ratio technique \cite{MBnubar10},  the CL
critical values for the fitting statistic $\Delta\chi^2 =\chi^2(point)-\chi^2(best)$  
as a function of the oscillation parameters $\Delta m^2$ and $\sin^22\theta$ are determined
from frequentist fake data studies.
Fig.~\ref{figure9} shows the MiniBooNE contours for
$\nu_e$ and $\bar \nu_e$ appearance oscillations in the
$200<E_\nu^{QE}<3000$~MeV energy range. 
The data indicate an oscillation signal region at greater than
95\% CL (99\% CL) with respect to a no-oscillation hypothesis for neutrino
(antineutrino) mode.  In neutrino mode, the MiniBooNE favored region
is somewhat below the LSND allowed region, but in antineutrino mode 
the MiniBooNE region is consistent with large parts of the LSND 99\% CL allowed
region and consistent with the limits from the KARMEN experiment \cite{Karmen2}.  

\section{Data Sets for Global Fits at High $\Delta m^2$ \label{global}}

The LSND and MiniBooNE appearance results 
must be considered within the context of
other relevant oscillation limits and signals.   Indeed, both LSND and MiniBooNE
data provide disappearance limits that are important constraints for
global fits.
Here we consider these disappearance results, as well as results from other SBL experiments.

\subsection{LSND and MiniBooNE Disappearance Searches \label{disresults}}

Both the LSND and the MiniBooNE data sets can also be used for disappearance
searches, in which the neutrinos oscillate to flavors that are not
observed in the detector.   In such a search, the highest precision
is achieved with
a ``near detector'' that constrains the unoscillated flux.
Therefore, for these analyses, LSND and MiniBooNE had to be paired
with sister experiments at closer distances--KARMEN
\cite{Karmen2} and SciBooNE \cite{MB_SB}, respectively. 

Both the LSND and KARMEN experiments made accurate measurements of 
the $\nu_e$-carbon cross section in the range $20<E_e<60$ MeV using
the reaction
$\nu_e + ^{12}C \rightarrow ^{12}N_{gs} + e^-$ \cite{LSNDxsec,KARMENxsec}.       
The cross section is measured by dividing 
the event rate by the predicted 
DAR flux, assuming no oscillations and with appropriate normalizations.    
Since KARMEN and LSND were at different distances from their neutrino sources, 
$\nu_e$ disappearance oscillations can induce a difference in the  measured cross
sections.  By comparing the measured cross sections,
accounting for normalization uncertainties \cite{BurmanISIS,BurmanLAMPF} 
and using the respective $L$ values, restrictions on the
allowed oscillation parameters \cite{ConradShaevitz} are obtained.

MiniBooNE has also performed 
$\nu_\mu$ and $\bar\nu_\mu$ disappearance searches  \cite{MBnudisapp,MB_SB}.
The strongest constraint comes from combining
MiniBooNE data with that from the SciBooNE detector, located 
a distance of 100 m from the neutrino source and having a 10.6-ton 
fiducial volume.  SciBooNE was a scintillator tracking detector as opposed
to the MiniBooNE mineral oil Cherenkov detector, but the
neutrino and antineutrino cross sections, as well
as the neutrino and antineutrino fluxes, are quite similar. 
The combined SciBooNE-MiniBooNE  data have been used to set
the world's best $\bar \nu_\mu$ disappearance limits for 
$\Delta m^2 < 20$ eV$^2$ \cite{MB_SB}.

\subsection{Summary of Short Baseline Data Sets  \label{others}}

\begin{figure}[tbp]
\begin{center}
\includegraphics[width=5.25in]{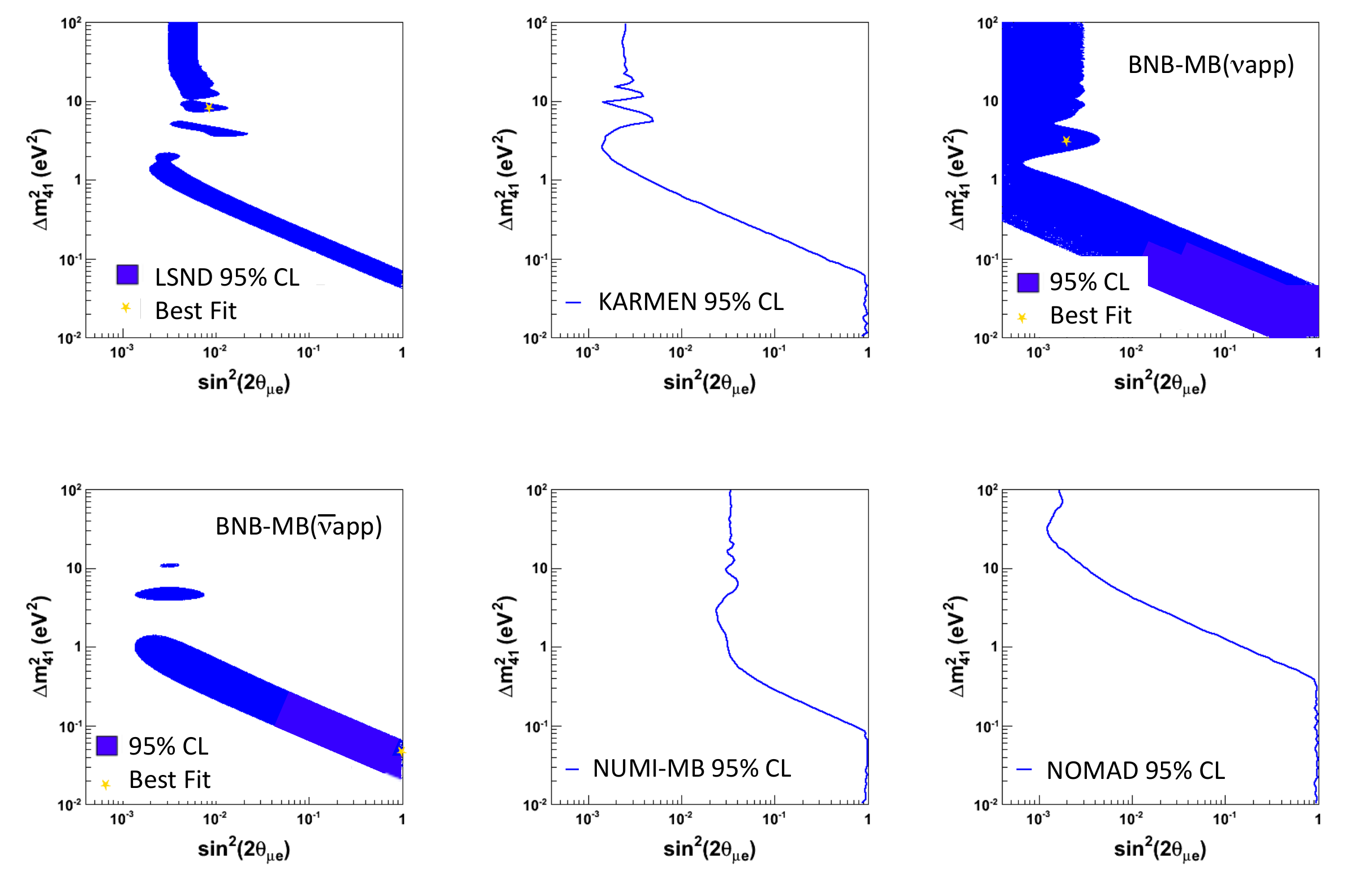}
\end{center}
\vspace{-0.5cm}
\caption{Summary of $\bar \nu_\mu \rightarrow \bar \nu_e$ and $
  \nu_\mu \rightarrow \nu_e$ results at 95\% C.L.  Top row:
  LSND \cite{lsndPRD}, KARMEN \cite{Karmen2}, MiniBooNE with BNB beam,
  $\nu$ \cite{MBnu};  Bottom row:
  MiniBooNE with BNB beam, $\bar{\nu}$ \cite{MBnubar},  MiniBooNE with
  NuMI beam \cite{MBNumi}, 
NOMAD \cite{NOMAD}.  From Ref. \cite{Ignarra}.
\label{figure10}}
\end{figure}

\begin{figure}[tbp]
\begin{center}
\includegraphics[width=5.25in]{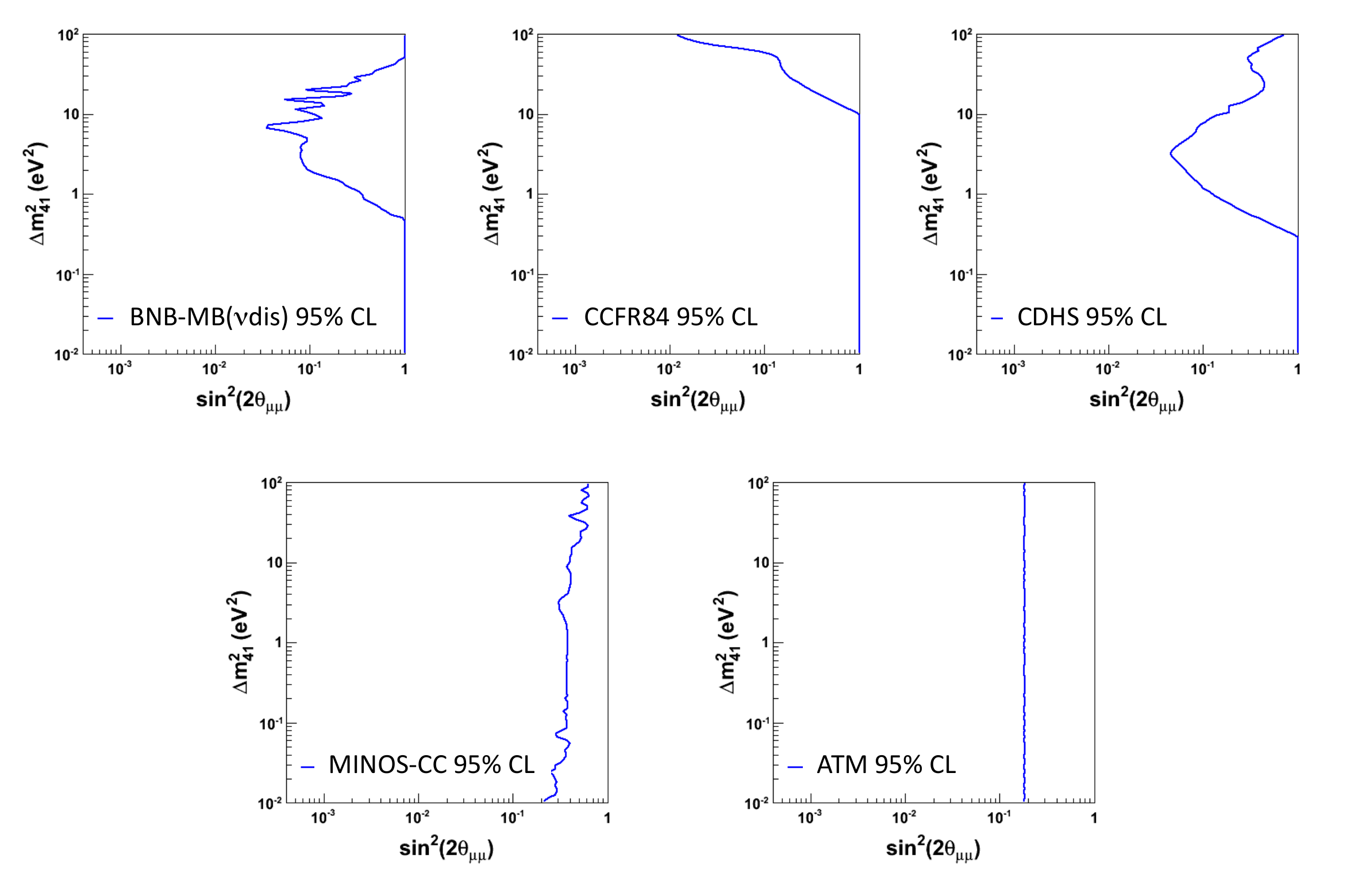}
\end{center}
\vspace{-0.5cm}
\caption{Summary of $\bar \nu_\mu \rightarrow \bar \nu_\mu$ and $
  \nu_\mu \rightarrow \nu_\mu$ results at 95\% C.L.  Top row:
MiniBooNE $\nu$ \cite{MBnudis},  CCFR \cite{CCFR84},  CDHS \cite{CDHS}; 
Bottom row: MINOS Charge Current data set \cite{MINOSdis}, Atmospheric
\cite{atmos}. From Ref. \cite{Ignarra}.
\label{figure11}}
\end{figure}

\begin{figure}[tbp]
\begin{center}
\includegraphics[width=5.25in]{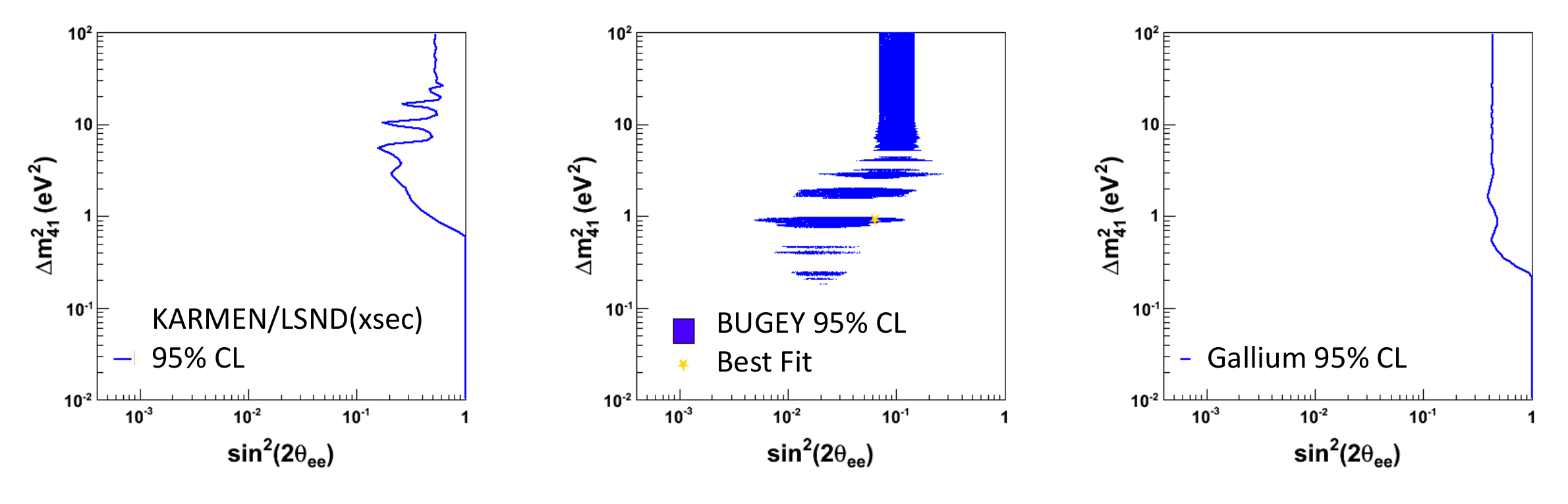}
\end{center}
\vspace{-0.5cm}
\caption{Summary of $\bar \nu_e\rightarrow \bar \nu_e$ and $
  \nu_e \rightarrow \nu_e$ results at 95\% C.L.   From left:
  KARMEN/LSND $\nu_e$ cross-section fit \cite{ConradShaevitz},  Bugey
  (and other reactor experiments) \cite{reactor}, and Gallium \cite{Gallium}. From Ref. \cite{Ignarra}.
\label{figure12}}
\end{figure}

In the next section, we will incorporate all data 
relevant to the high $\Delta m^2$ region into a single model. 
Figures~\ref{figure10}, \ref{figure11}, and \ref{figure12} show the data
sets involved in the fits, presented as 
individual two-neutrino oscillation fits (Eqs.~\ref{osceq} and  
\ref{disosceq}) \cite{Ignarra}.   The data are categorized as $\nu_\mu
\rightarrow \nu_e$ appearance
searches, $\nu_\mu$ disappearance searches, and $\nu_e$ disappearance
searches.   Signals show up at 95\% CL in the LSND, MiniBooNE,
and Bugey reactor experiment \cite{reactor} data sets. The reactor
signal is a recent observation based on a reanalysis of the reactor
flux \cite{reactor}.   The other data sets have no closed contour at 95\%
CL, and so a limit is shown.   However, it should be noted that the  KARMEN/LSND
cross section, discussed above in Sec.~\ref{disresults}, and the Gallium data sets \cite{Gallium}, from
calibration data of SAGE  \cite{SAGE3}
and GALLEX \cite{GALLEX3}, both present closed
contours at 90\% CL.  

Two issues for future analyses should be noted.
First, if the Gallium data is corrected for recent
estimates of the cross section \cite{Giuntixsec}, then the result
would show a closed contour at 95\% CL.   Second, the MiniBooNE disappearance limits used in the
global fits pre-date the final MiniBooNE-SciBooNE
analyses \cite{MBnudisapp,MB_SB},  and are slightly less stringent.
However, we do not expect either of these issues to change significantly the
overall conclusions of the next section.

\section{LSND and MiniBooNE within the Context of Global Fits \label{sterilecon}}

We now present the LSND and MiniBooNE results within the context of
global fits involving sterile neutrinos, following the phenomenology
of Sec.~\ref{pheno}.
As described in Ref.~\cite{Ignarra}, global fits are
derived from Markov chain-based scans \cite{Markov} 
from 0.01 eV$^2$ to 100 eV$^2$.  Systematic and
statistical errors are included.   

We will quantify the quality of the fits through
the $\chi^2$/dof and the compatibility of subsets through the
Parameter Goodness-of-Fit (PGF) test \cite{pgtest}.  
We will use two specific cases of the PGF test, dividing the data 
into appearance {\it vs.} disappearance data sets and neutrino {\it
  vs.} antineutrino data sets, using the respective definitions:
\begin{equation}
R_{PGF}^{appdis}=  (\chi^2_{min,global} - \chi^2_{min,app}
-\chi^2_{min,dis})/(N_{app}  + N_{dis} - N_{global})   \label{PGFapp}
\end{equation}
and 
\begin{equation}
R_{PGF}^{\nu \bar \nu}= (\chi^2_{min,global} - \chi^2_{min,\nu}
-\chi^2_{min,\bar\nu})/(N_{\nu}  + N_{\bar \nu} - N_{global}).   \label{PGFnu}
\end{equation}
In the above, the numerator is a function of the minimum $\chi^2$ of the global fit
and the subsets, while the denominator 
is a function of the number of independent parameters, $N$, in the corresponding
fit.     If the global best fit parameters are
similar to those from the subset fits, then the $\chi^2_{PGF}$ 
value will be small and will indicate
good compatibility when the probability of $R_{PGF}$ is evaluated
as a $\chi^2/dof$. 

\subsection{The Problem with 3+1 Fits \label{threeplusone}}

Referring to Sec.~\ref{pheno}, 
a 3+1 fit has three parameters:  $\Delta m^2_{41}$, $|U_{e4}|$ and
$|U_{\mu 4}|$. The two matrix elements are related to the
mixing angles according to Eqs.~\ref{angleapp}, \ref{anglemumu}, and
\ref{angleee}.

Before fitting all of the data sets, a simple calculation can be used to show 
that the combined appearance and disappearance results from LSND and MiniBooNE
alone  already stress this model in the $\Delta m^2 > 1$ eV$^2$ region.  
In a 3+1 model, the mixing angle limit from the LSND-KARMEN cross
section analysis translates to a limit on
$|U_{e 4}^2|$, through Eq.~\ref{angleee}, that is roughly
$\lesssim 0.05$, although there are large variations
with $\Delta m^2$.   The stringent $\sin^22\theta_{\mu \mu}$  limit
from the MiniBooNE-SciBooNE joint analysis corresponds to  $|U_{\mu
  4}|^2<0.025$,
using Eq.~\ref{anglemumu}.  Thus, the disappearance
results favor a very small appearance mixing angle, which, from  
Eq.~\ref{angleapp}, is about $\sin^2 2\theta_{\mu e}\sim 0.005$ and
is not in good agreement with LSND and MiniBooNE.    Therefore, the LSND and
MiniBooNE data alone will force a lower $\Delta m^2$ solution.  

This is consistent with what is seen in the global fit \cite{Ignarra}, which 
yields a   $\chi^2_{min}/dof$
of 233.9/237 with a 55\% probability for this best fit point
and a $\chi^2_{null}/dof$ of 286.5/240 with a 2.1\% probability.
The best fit parameters are  0.92 eV$^2$,  0.17, and 0.15, for   
$\Delta m^2$, $|U_{e4}|$, and   $|U_{\mu 4}|$, respectively.  Consequently, the
$\Delta m^2$ of this solution sits just below 1 eV$^2$.

In contrast, the compatibility for this 3+1 model between 
appearance and disappearance (from Eq.~\ref{PGFapp}) 
is found to be only 0.0013\%, and the
compatibility between $\nu$ and $\overline{\nu}$ (from
Eq.~\ref{PGFnu}) is 0.14\%.   These
very poor compatibilities are a warning that some data sets
have best fit parameters in conflict with that found in the
global fit.    As a result, one is led to conclude that
 3+1 models are, at best, marginal descriptions of the data.

\subsection{Potential Success of 3+2 Global Fits  \label{threeplustwo}}

The poor compatibility of the data sets can possibly be improved
by expanding to a 3+2 model, which  
introduces four new parameters:  another high mass eigenstate,
two additional mixing parameters, and a $CP$ phase.    The latter
allows for appearance signals for neutrinos that differ from
antineutrinos.    This extra degree of freedom significantly improves
the compatibility between neutrino and antineutrino data sets, but it
does not address the conflict between appearance and disappearance
data sets.

From Ref.~\cite{Ignarra}, the best fit parameters for a 3+2 model are:
0.92 eV$^2$, 17 eV$^2$, 0.13, 0.15,  0.16,  0.069, and 1.8$\pi$,
for the parameters 
$\Delta m^2_{41}$, $\Delta m^2_{51}$, $|U_{\mu4}|$, $|U_{e4}|$, 
$|U_{\mu5}|$, $|U_{e5}|$, and
$\phi_{54}$  (the $CP$ phase), respectively.       The results give
good $\chi^2$ probabilities for a signal, with a $\chi^2_{min}/dof$
of 221.5/233 and a 69\% probability for the best fit point
and a $\chi^2_{null}/dof$ of 286.5/240 with a 2.1\% 
probability--very similar to the 3+1 results.  
The PGF for the $\nu$ versus $\overline{\nu}$
data set comparison rises to 5.3\%, which is acceptable.

\begin{figure}[tbp]
\begin{center}
\includegraphics[width=5.75in]{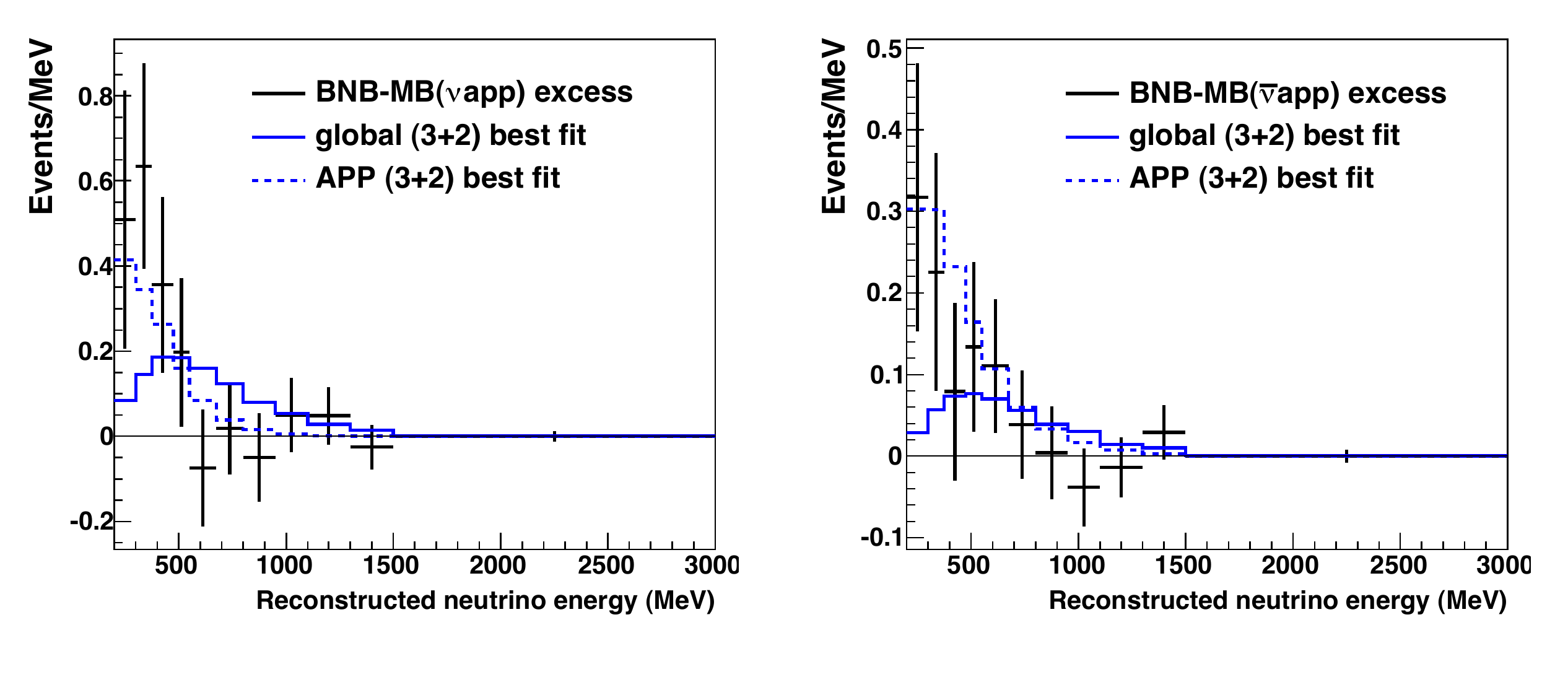}
\end{center}
\vspace{-0.5cm}
\caption{Illustration of the problem presented by the MiniBooNE data
  in global 3+2 fits to oscillations.  Left, excess in neutrino mode;
  Right, excess in antineutrino mode.   Solid line is global best fit;
  Dashed line is a 3+2 fit to only the appearance
  data (Plot from Ref.~\cite{Ignarra}).
\label{figure13}}
\end{figure}

\begin{figure}[tbp]
\begin{center}
\includegraphics[width=5.75in, clip=T]{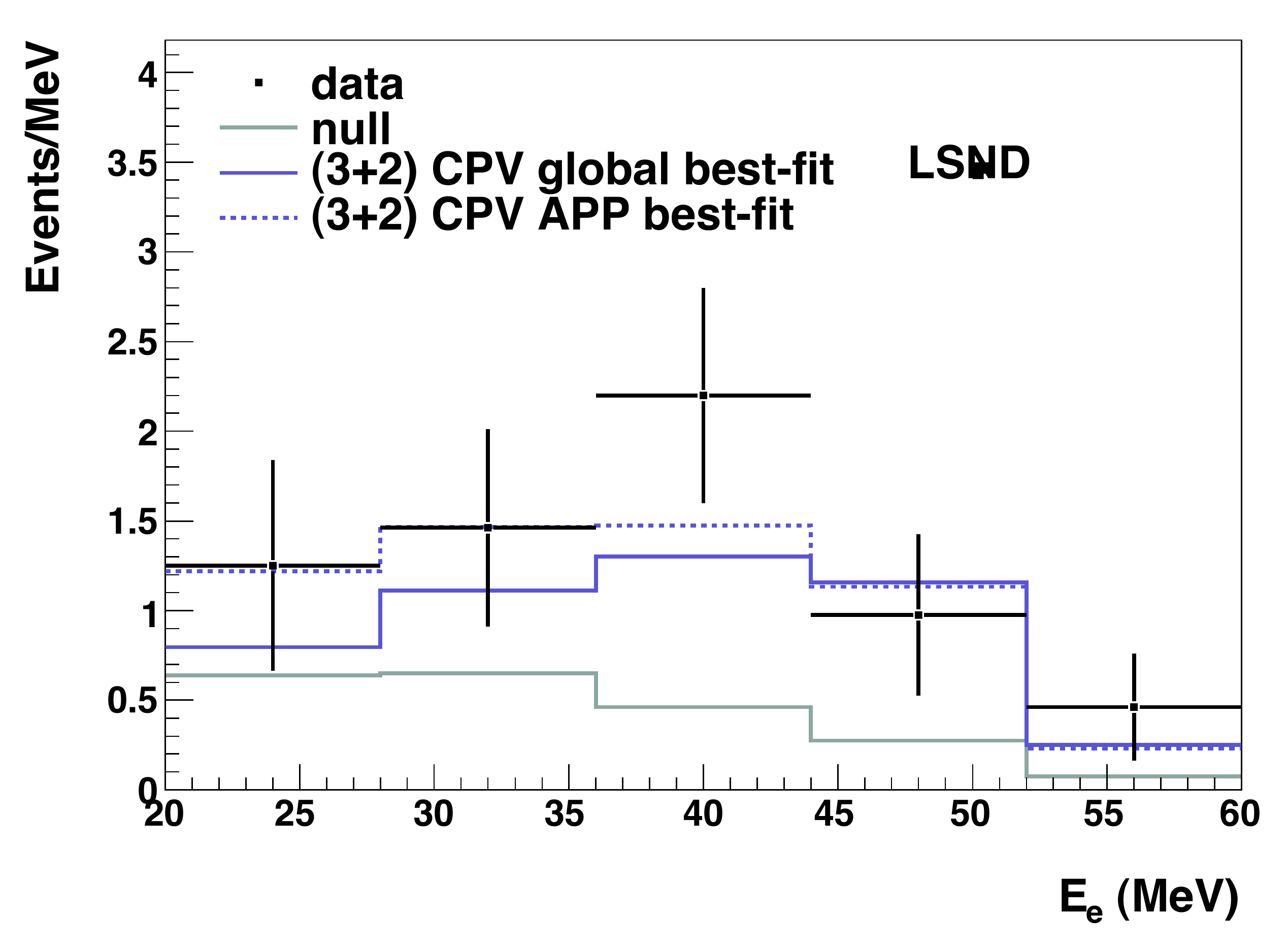}
\end{center}
\vspace{-0.5cm}
\caption{Comparison to the LSND data set of the global 3+2 fit to oscillations
and the appearance-only fit.
\label{figure14}}
\end{figure}

It is striking, however,  that the PGF for the appearance versus disappearance
data sets slightly worsens from the 3+1 case to 0.0082\%.
The source of the issue can be tracked to the MiniBooNE low-energy
excess.   The fit to the  $\nu$ and $\bar \nu$ appearance signals alone
are internally consistent, assuming a non-zero $CP$ phase, but the
best fit
is strikingly different from the global fit.    This can be seen
clearly in Fig.~\ref{figure13}, 
where the MiniBooNE electron-like excess is shown for both neutrino and
antineutrino modes and two example 3+2 fits are
overlaid.   The solid lines show the expectations for the global best
fit.  The dashed lines show the best fits to only the appearance data
sets.  The dashed lines indicate a good representation of
the data, while the global fit cannot explain the rise in events at
low energy.   In fact, the
parameter
set for the appearance-only best fit is excluded by the disappearance data.  
Fig.~\ref{figure14} shows that, in contrast to MiniBooNE, the LSND appearance data set is in
agreement with the 3+2 global fit.    The normalization of LSND is
approximately 30\% higher than the global fit, but the 
energy-dependent shape is well described.  Therefore, the poor PGF has been 
interpreted as indicating an issue with the MiniBooNE data in the
global fits. For example, it has been suggested that multi-nucleon nuclear 
effects could cause the neutrino energy to be underestimated for some
fraction of the events \cite{multi-nucleon}.

Rather than indicating a problem with the MiniBooNE data, however,
this may instead be pointing to a limitation of the PGF description of compatibility.   
Consider Fig.~\ref{figure13} once again.   
After subtracting the  3+2 best fit prediction 
from the low-energy excess, the residual excess for neutrinos and
antineutrinos can be fit to an enhanced $\pi^0$ 
background model.  A 20\%  increase of the $\pi^0$  background 
normalization gives a $\chi^2/dof$ for the residual excess of
17.8/19.  The systematic error on the MiniBooNE 
$\pi^0$ background is 14.5\% (13.9\% ) in 
neutrino (antineutrino) mode, and, therefore,
this 20\% discrepancy is only at the 
1.4$\sigma$ level.    This is not a particularly large deviation, given the
number of data sets in the global fit, and introduces concerns
about the validity of the PGF compatibility statistic.     

This apparent discrepancy may indicate that the PGF 
has difficulty properly characterizing compatibility when systematic 
uncertainties can mimic a signal.
The reason that the PGF is returning a poor
compatibility can be seen from considering the numerator in
Eq.~\ref{PGFapp}.      When appearance data are fit alone, a smaller
$\chi^2$ is found when the entire MiniBooNE low-energy
excess is attributed to an oscillation signal than when the global fit
parameters plus a 1.4$\sigma$ fluctuation of background are evaluated.
This combination yields a large value for  $R_{PGF}^{appdis}$, and hence a poor
PGF, even though a compatible solution was available.    

This 3+2 model can be tested in the near future, 
as it makes specific predictions for the MicroBooNE
experiment \cite{microboone}, which will begin running in late 2013.
The MicroBooNE experiment will use a 
liquid argon time projection chamber (LArTPC) 60 ton fiducial-volume, 
which will be located on the BNB
close to the MiniBooNE detector.
The  LArTPC can differentiate events which 
have a final state electron from those with a final state photon.
Unlike MiniBooNE, with LArTPC selection cuts, MicroBooNE
will have very little contribution from photon-producing background sources, such
as misidentified $\pi^0$ events.   Thus, MicroBooNE will definitively
show whether the MiniBooNE low-energy excess in neutrino mode is
associated with electron neutrino events possibly from oscillations
or events with photons from some
other process.

\section{Conclusions \label{thatsallfolks}}

Neutrino oscillations have been an unexpected and rich area for
particle physics studies
over the past several decades.  The conclusive observations 
that neutrinos have mass and that lepton flavor
is not conserved have changed the direction of the field.  Now, the indications
from LSND, 
MiniBooNE, and other experiments that a new type of sterile neutrino might exist
could have an equally important impact.  Both LSND and MiniBooNE see 
indications of electron neutrino/antineutrino appearance, but there are discrepancies
in the energy distributions of the appearance signal that could or could not be 
related to oscillations.  The analysis of all high $\Delta m^2$ data
sets shows that models with several sterile neutrinos can give
acceptable global fits.   Reported incompatibility with sterile neutrino models, 
especially between the appearance and disappearance
measurements, may be arising from the choice of test statistic, rather
than an underlying discrepancy with the data sets. 
Investigations of high $\Delta m^2$ oscillations currently form one of 
the most active areas of neutrino physics and many new experiments are being mounted
or considered.   This future program builds on the initial measurements
of LSND and MiniBooNE discussed here and aims to 
provide a definitive  exploration of 
new physics signals in the neutrino sector, such as sterile neutrinos.

\section{Acknowledgments}
This work was made possible by the dedicated efforts of the LSND 
and MiniBooNE collaborations.   We thank C. Ignarra, G. Karagiorgi and
Z. Pavlovic for their input.   JMC and MHS thank the National Science
Foundation for support.   WL thanks the Department of Energy for support.

~

\begin{thebibliography}{9}

\bibitem{1209.3023}    M.~C.~Gonzalez-Garcia, M.~Maltoni, J.~Salvado and T.~Schwetz,
  arXiv:1209.3023 [hep-ph].

\bibitem{whitepaper}
K.~N.~Abazajian {\it et al.},
arXiv:1204.5379 [hep-ph].


\bibitem{Zdecay} J.~Beringer {\it et al.}  [Particle Data Group],
Phys.\ Rev.\  D {\bf 86}, 010001 (2012).


\bibitem{lsndPRL95} 
  C.~Athanassopoulos {\em et~al.},
  Phys.\ Rev.\ Lett. 75, 2650 (1995)


\bibitem{lsndPRL96}
  C.~Athanassopoulos {\em et~al.},
  Phys.\ Rev.\ Lett. 
  77, 3082 (1996).
 
\bibitem{lsndPRL98}
  C.~Athanassopoulos {\em et~al.},
  Phys.\ Rev.\ Lett. 
  81, 1774 (1998).
 
\bibitem{lsndPRD}
  A.~Aguilar {\em et~al.},
  Phys.\ Rev.\ D 64, 112007 (2001).

\bibitem{MBprop}  
E.~Church {\it et al.},
``A proposal for an experiment to measure $\nu_\mu \rightarrow \nu_e$
oscillations and $\nu_\mu$ disappearance at the Fermilab Booster: BooNE'',
submitted to the Fermilab Program Advisory Committee (1997).

\bibitem{MBnu}    A.~A.~Aguilar-Arevalo {\it et al.}  [MiniBooNE Collaboration],
  [arXiv:0704.1500 [hep-ex]].
  
  \bibitem{MBnubar09}    A.~A.~Aguilar-Arevalo {\it et al.}  [MiniBooNE Collaboration],
  Phys.\ Rev.\ Lett.\  {\bf 103}, 111801 (2009)
  [arXiv:0904.1958 [hep-ex]].
  
  \bibitem{MBnubar}   A.~A.~Aguilar-Arevalo {\it et al.}  [MiniBooNE Collaboration],
  arXiv:1207.4809 [hep-ex].

 
 \bibitem{sorel}  M.~Sorel, J.~M.~Conrad and M.~Shaevitz,
  Phys.\ Rev.\  D {\bf 70}, 073004 (2004).

\bibitem{georgiaCP}    G.~Karagiorgi, A.~Aguilar-Arevalo, J.~M.~Conrad, M.~H.~Shaevitz, K.~Whisnant, M.~Sorel and V.~Barger,
  Phys.\ Rev.\ D {\bf 75}, 013011 (2007)
  [Erratum-ibid.\ D {\bf 80}, 099902 (2009)]
  [hep-ph/0609177].

\bibitem{georgiaVaiability}  G.~Karagiorgi, Z.~Djurcic, J.~M.~Conrad, M.~H.~Shaevitz and M.~Sorel,
  Phys.\ Rev.\  D {\bf 80}, 073001 (2009)
  [Erratum-ibid.\  D {\bf 81}, 039902 (2010)].

\bibitem{Ignarra}   J.~M.~Conrad, C.~M.~Ignarra, G.~Karagiorgi, M.~H.~Shaevitz and J.~Spitz,
  arXiv:1207.4765 [hep-ex].

\bibitem{DB}    F.~P.~An {\it et al.}  [Daya Bay Collaboration],
Phys.\ Rev.\ Lett.\ {\bf 108}, 171803 (2012);
  arXiv:1210.6327 [hep-ex].

\bibitem{DCGd}
  Y.~Abe {\it et al.}  [Double Chooz Collaboration],
  Phys.\ Rev.\ D {\bf 86}, 052008 (2012)
  [arXiv:1207.6632 [hep-ex]].


\bibitem{KL}    S.~Abe {\it et al.}  [KamLAND Collaboration],
  Phys.\ Rev.\ Lett.\  {\bf 100}, 221803 (2008)
  [arXiv:0801.4589 [hep-ex]].

\bibitem{MINOS}
  P.~Adamson {\it et al.}  [MINOS Collaboration],
  Phys.\ Rev.\ Lett.\  {\bf 107}, 181802 (2011)
  [arXiv:1108.0015 [hep-ex]].

\bibitem{MINOSdis}   P.~Adamson {\it et al.}  [MINOS Collaboration],
  Phys.\ Rev.\ Lett.\  {\bf 108}, 191801 (2012)
  [arXiv:1202.2772 [hep-ex]].

\bibitem{RENO}    J.~K.~Ahn {\it et al.}  [RENO Collaboration],
  Phys.\ Rev.\ Lett.\  {\bf 108}, 191802 (2012)
  [arXiv:1204.0626 [hep-ex]].


\bibitem{SK}  
  R.~Wendell {\it et al.}  [Super-Kamiokande Collaboration],
  Phys.\ Rev.\ D {\bf 81}, 092004 (2010)
  [arXiv:1002.3471 [hep-ex]].


\bibitem{SNOfinal} 
  B.~Aharmim {\it et al.}  [SNO Collaboration],
  arXiv:1109.0763 [nucl-ex].


\bibitem{T2K}   K.~Abe {\it et al.}  [T2K Collaboration],
  Phys.\ Rev.\ Lett.\  {\bf 107}, 041801 (2011)
  [arXiv:1106.2822 [hep-ex]].

\bibitem{T2Kdis}   K.~Abe {\it et al.}  [T2K Collaboration],
  Phys.\ Rev.\ D {\bf 85}, 031103 (2012)
  [arXiv:1201.1386 [hep-ex]].

\bibitem{bib:burman}
R.\ L.\ Burman, M.\ E.\ Potter, and E.\ S.\ Smith,
\emph{Nucl.\ Instrum.\ Methods A} {\bf 291}, 621, (1990);
R.\ L.\ Burman, A.\ C.\ Dodd, and P.\ Plischke,
\emph{Nucl.\ Instrum.\ Methods in Phys.\ Research A} {\bf 368}, 416, (1996).

\bibitem{lsndNIM}    C.~Athanassopoulos {\it et al.}  [LSND Collaboration],
  Nucl.\ Instrum.\ Meth.\ A {\bf 388}, 149 (1997)
  [nucl-ex/9605002].

\bibitem{veto}
J.\ J.\ Napolitano {\it et.\ al.\ },
\emph{Nucl.\ Instrum.\ Methods A} {\bf 274}, 152, (1989).


\bibitem{mb_flux}
A.~A.~Aguilar-Arevalo {\em et~al.},
Phys.\ Rev.\ D 79, 072002 (2009).

\bibitem{mb_detector}
  A.~A.~Aguilar-Arevalo {\em et~al.},
  Nucl.\ Instrum.\ Meth.\  A 599, 28 (2009).

\bibitem{SamAnnualReview} H. Gallagher, G. Garvey, and G.P. Zeller,
Annu. Rev. Nucl. Part. Sci. {\bf 61}, 355 (2011)

\bibitem{numubkgd}
This background also includes contributions from $\bar \nu_\mu C
\rightarrow \mu^+ n X$ and $\nu_\mu C \rightarrow \mu^- n X$.

\bibitem{Kardet}  G. Drexlin, {\it et al.}, 
Nucl. Instr. and Meth., A {\bf 289} 490 (1990).

\bibitem{Karmen2}    B.~Armbruster {\it et al.}  [KARMEN Collaboration],
  Phys.\ Rev.\ D {\bf 65}, 112001 (2002)
  [hep-ex/0203021].

\bibitem{ref:joint} E. D. Church, K. Eitel, G. B. Mills, and
M. Steidl, \emph{Phys. Rev. D}{\bf 66}, 013001, (2002).

\bibitem{unexplained}    A.~A.~Aguilar-Arevalo {\it et al.}  [MiniBooNE Collab
oration],
  [arXiv:0812.2243 [hep-ex]].  

  \bibitem{MBnubar10} A.~A.~Aguilar-Arevalo {\it et al.}  [MiniBooNE Collaboration],
  Phys.\ Rev.\ Lett.\  {\bf 105}, 181801 (2010)
  [arXiv:1007.1150 [hep-ex]].

\bibitem{ChengKProd}
G.~Cheng {\it et al.}  [SciBooNE Collaboration],
  Phys.\ Rev.\ D {\bf 84}, 012009 (2011)
  [arXiv:1105.2871 [hep-ex]],
   C.~Mariani, G.~Cheng, J.~M.~Conrad and M.~H.~Shaevitz,
  Phys.\ Rev.\ D {\bf 84}, 114021 (2011)
  [arXiv:1110.0417 [hep-ex]].
  
  \bibitem{pgtest}   M.~Maltoni and T.~Schwetz,
  Phys.\ Rev.\ D {\bf 68}, 033020 (2003)
  [hep-ph/0304176].

\bibitem{MB_SB}
G.~Cheng {\em et~al.},
arXiv:1208.0322 [hep-ex] (2012).

\bibitem{LSNDxsec}
  L.~B.~Auerbach {\it et al.}  [LSND Collaboration],
  Phys.\ Rev.\  C {\bf 64}, 065501 (2001).

\bibitem{KARMENxsec}   B.~E.~Bodmann {\it et al.}  [KARMEN Collaboration.],
  Phys.\ Lett.\ B {\bf 332}, 251 (1994). B.~Armbruster, {\it et al.}
  [KARMEN Collaboration], Phys. Rev. C57, 3414 (1998).

\bibitem{BurmanISIS}  R.~L.~Burman,
  Nucl.\ Instrum.\ Meth.\  A {\bf 368}, 416 (1996).

\bibitem{BurmanLAMPF}  R.~L.~Burman, M.~E.~Potter and E.~S.~Smith,
  Nucl.\ Instrum.\ Meth.\  A {\bf 291}, 621 (1990).

\bibitem{ConradShaevitz}   J.~M.~Conrad and M.~H.~Shaevitz,
  Phys.\ Rev.\ D {\bf 85}, 013017 (2012)
  [arXiv:1106.5552 [hep-ex]].

\bibitem{MBnudisapp}
K.~B.~M.~Mahn {\it et al.}  [SciBooNE and MiniBooNE Collaboration],
  Phys.\ Rev.\ D {\bf 85}, 032007 (2012)
  [arXiv:1106.5685 [hep-ex]].

 \bibitem{Markov}
   S.~Hannestad,
  arXiv:0710.1952 [hep-ph].

\bibitem{reactor}     G.~Mention, M.~Fechner, T.~.Lasserre, T.~.A.~Mueller, D.~Lhuillier, M.~Cribier and A.~Letourneau,
  Phys.\ Rev.\ D {\bf 83}, 073006 (2011)
  [arXiv:1101.2755 [hep-ex]].

\bibitem{Gallium}    M.~A.~Acero, C.~Giunti and M.~Laveder,
  Phys.\ Rev.\ D {\bf 78}, 073009 (2008)
  [arXiv:0711.4222 [hep-ph]].


\bibitem{SAGE3}   J.~N.~Abdurashitov {\it et al.}  [SAGE Collaboration],
  Phys.\ Rev.\  C {\bf 80}, 015807 (2009).

\bibitem{GALLEX3}   F.~Kaether, W.~Hampel, G.~Heusser, J.~Kiko and T.~Kirsten,
  Phys.\ Lett.\  B {\bf 685}, 47 (2010).


\bibitem{Giuntixsec}    C.~Giunti and M.~Laveder,
  Phys.\ Rev.\ C {\bf 83}, 065504 (2011)
  [arXiv:1006.3244 [hep-ph]].

\bibitem{MBNumi}
  P.~Adamson {\it et al.},
  Phys.\ Rev.\ Lett.\  {\bf 102}, 211801 (2009)
  [arXiv:hep-ex/0809.2447].

\bibitem{NOMAD}
P.~Astier {\it et al.}  [NOMAD Collaboration],
  Phys.\ Lett.\ B {\bf 570}, 19 (2003)
  [arXiv:hep-ex/0306037];
%
  D.~Gibin,
  Nucl.\ Phys.\ Proc.\ Suppl.\  {\bf 66}, 366 (1998);
  V.~Valuev  [NOMAD Collaboration],
  {\it Prepared for International Europhysics Conference on High-Energy Physics (HEP 2001), Budapest, Hungary, 12-18 Jul 2001}.
  
\bibitem{MBnudis}
G.~Cheng {\it et al.},
Phys.\ Rev.\ D {\bf 86}, 052009 (2012)
[arXiv:hep-ex/1208.0322].

\bibitem{CCFR84}
I.~Stockdale {\it et al.} [CCFR Collaboration],
Phys.\ Rev.\ Lett.\ 52, 1384 (1984),
Z. Phys. C 27, 53 (1985).

\bibitem{CDHS}    F.~Dydak {\it et al.},
  Phys.\ Lett.\ B {\bf 134}, 281 (1984).

\bibitem{atmos}
 M.~Maltoni, T.~Schwetz, M.~A.~Tortola and J.~W.~F.~Valle,
  New J.\ Phys.\  {\bf 6}, 122 (2004)
  [hep-ph/0405172].

\bibitem{multi-nucleon}
M.~Martini, M.~Ericson, and G.~Chanfray,
arXiv:1202.4745 [hep-ph] (2012); 
O.~Lalakulich and U.~Mosel, 
arXiv:1208.3678 [nucl-th] (2012):
J.~Nieves, F.~Sanchez, I.~Ruiz Simo, and M.~J.~Vicente Vacas, 
arXiv:1204.5404 [hep-ph] (2012).

\bibitem{microboone}
G.~Karagiorgi [MicroBooNE Collaboration],
  J.\ Phys.\ Conf.\ Ser.\  {\bf 375}, 042067 (2012).

\end{thebibliography}
\end{document}